# Normalization, orthogonality and completeness of quasinormal modes of open systems: the case of electromagnetism

C. Sauvan[1], T. Wu[2], R. Zarouf[3], E. A. Muljarov[4], P. Lalanne[2*]

[1]Université Paris-Saclay, Institut d'Optique Graduate School, CNRS, Laboratoire Charles Fabry, 91127 Palaiseau, France

[2]Laboratoire Photonique, Numérique et Nanosciences (LP2N), IOGS-University of Bordeaux-CNRS, 33400 Talence cedex, France

[3]Aix-Marseille Université, Laboratoire ADEF, Campus Universitaire de Saint-Jérôme, 52 Avenue Escadrille Normandie Niemen, 13013 Marseille, France

[4]School of Physics and Astronomy, Cardiff University, Cardiff CF24 3AA, United Kingdom

*philippe.lalanne@institutoptique.fr

**Abstract:**

The scattering of electromagnetic waves by resonant systems is determined by the excitation of quasinormal modes (QNMs), i.e., the eigenmodes of the system. This Review addresses three fundamental concepts in relation with the representation of the scattered field as a superposition of the excited QNMs: normalization, orthogonality, and completeness. Orthogonality and normalization enable a straightforward assessment of the QNM excitation strength for any incident wave. Completeness guaranties that the scattered field can be faithfully expanded into the QNM basis. These concepts are non trivial for non-conservative (non-Hermitian) systems and have driven many theoretical developments since initial studies in the 70's. Yet, owing to recent achievements, they are not easily grasped from the extensive and scattered literature, especially for newcomers in the field. After recalling fundamental results obtained in early studies on the completeness of the QNM basis for simple resonant systems, we review recent achievements and debates on the normalization, clarify under which circumstances the QNM basis is complete, and highlight the concept of QNM regularization with complex coordinate transform.

## 1. Introduction

<u>1.1 Modes in conservative and non-conservative systems</u>

Modes are central in physics, since the most general evolution of a system is a superposition of its modes, or eigenstates. From textbooks, everyone knows the special case of conservative systems studied in the mathematical framework of Hilbert spaces, e.g., particles in a potential with infinite barriers in quantum mechanics, cavities with perfectly-conducting metallic walls in electromagnetics, non-dissipative vibrating ropes with a finite length, or photonic crystals made with a perfectly periodic arrangement of non-absorbing materials.

Conservative systems have simple imposed boundary conditions (Dirichlet, Neuman, or Born-von Karman conditions) and the operators describing them are Hermitian (or self-adjoint). The eigenstates of conservative systems are usually called normal modes. They form a complete set, while the corresponding eigenvalues (frequencies or energies) are real. Moreover, it is easy to define an inner product for normal modes, namely the usual scalar product of the corresponding Hilbert space. Many beautiful consequences immediately follow, such as the mode orthogonality or the well-developed perturbation theory for linear problems. For a didactical introduction to Hermitian or conservative electromagnetic systems, we refer the interested reader to Chapters 7 and 8 in [Har61] and Chapter 2 in [Joa08].

Conservative systems are obviously idealized systems. In reality, there exists flows (of energy, particle, information…) into external degrees of freedom that are outside the system subspace under consideration. In electromagnetism, there is always a given amount of loss, either by absorption or by radiative leakage in an open space. Such real open systems are referred to as non-Hermitian or non-conservative and are described by non-self-adjoint operators (non-Hermitian matrices).

The properties of non-conservative systems are much richer than those of conservative ones. The other side of the coin is that the mathematical tools used for their conceptualization are more difficult [Lal18]. The nice properties of Hilbert spaces are lost and the eigenstates are no longer normal modes: eigenvalues are now complex and sometimes unstable regarding small perturbations of the system [Jar21], the modes form a complete set only under certain conditions, and the usual inner product of Hermitian systems is unfortunately invalid. The latter cannot be defined because the field of the eigenstates exponentially diverges at long distance. Therefore, non-Hermitian systems require more mathematical rigor and heuristics should be considered with care. However, it is still possible to prove that the eigenstates are orthonormal with a suitable regularization of the diverging fields and a redefinition of the inner product. Defining a correct inner product that allows to recover the property of orthonormalization is extremely helpful to extend the well-known tools of conventional Hermitian electrodynamics to non-Hermitian systems.

Note that the notion of non-Hermitian systems spreads in physics well beyond the present electromagnetic context, e.g., in mechanics, gravitation, acoustics, or atomic, nuclear, and molecular physics. The field of non-Hermitian physics has garnered widespread interest in recent years in relation with prominent topics such as unidirectional invisibility, enhanced sensitivity, topological energy transfer, coherent perfect absorption, single-mode lasing, and robust biological transport, see [Ash20] for a recent and extensive review on this manifold topic.

## 1.2 Quasinormal modes, resonant states, leaky modes…

The interest in the eigenstates of non-conservative systems probably started in quantum mechanics with the calculation of alpha decay [Gam28] or the cross-section for nuclear reactions [Sie39]. However, since they are the solutions of a source-free wave equation, these states can be encountered in any field of wave physics. Over the years, different communities have given them various names, each one putting the emphasis on a different characteristic.

Quantum mechanics chose to underline their finite lifetime (unstable states [Zel61], decaying states [Mor71]), their link with the system resonances (resonant states [Mor73]), or the fact that they are the natural modes of the scatterer [Hoe79]. The term "quasinormal mode" was likely coined in the context of black holes and gravitational waves [Jar21,Pre71,Kok99,Kon11] before being used for optical cavities [Lai90,Leu94a]. In general relativity, the scattering potential has an infinite spatial extent and not a compact support. As a result, the modes do not form a complete set and that was the main reason to call them "quasi". In electromagnetism, the scattering potential (e.g., a permittivity change $\Delta\varepsilon$) has usually a compact support and instability problems are rarely encountered. In this context, the prefix "quasi" was chosen to highlight the differences but also the similarities between Hermitian systems, which support normal modes, and non-Hermitian systems, which support modes that can be seen as almost normal under a suitable regularization.

The eigenmodes of open systems are also known as leaky modes in optical waveguides theory [Sny83] and in acoustics [Hei04]. Finally, let us mention the name "morphology-dependent resonance" that is sometimes given to the optical modes of microparticles [Leu96,Lee99].

Hereafter, we refer to the eigenmodes of non-Hermitian open systems as quasinormal modes, abbreviated in QNMs.

## 1.3 Scope of the review

This review focuses on the normalization, the orthogonality, and the completeness of electromagnetic QNMs. For a review of the numerous and various applications of QNMs in photonics, we refer the interested reader to [Lal18,Kri20,Wu21a].

Let us recall that completeness and orthonormalization are essential properties in the construction of many theories. First, completeness underpins everything: the eigenstates $\boldsymbol{\Psi}_m$ constitute a relevant set to decompose scattered waves only if they form a complete set. One usually write that any scattered wave $\boldsymbol{\Psi}_s$ can be written as a superposition of modes, $\boldsymbol{\Psi}_s = \Sigma_m \alpha_m \boldsymbol{\Psi}_m$, where $\alpha_m$ is the amplitude (or excitation coefficient) of the $m^{\text{th}}$ mode. In classical textbooks on conservative systems, normalization is directly related to orthogonality through the concept of inner product. With the standard bra-ket notation, $\langle \boldsymbol{\Psi}_m | \boldsymbol{\Psi}_p \rangle$ is the inner product between eigenstates $\boldsymbol{\Psi}_m$ and $\boldsymbol{\Psi}_p$. The latter are orthogonal if $\langle \boldsymbol{\Psi}_m | \boldsymbol{\Psi}_p \rangle = 0$ and $\langle \boldsymbol{\Psi}_m | \boldsymbol{\Psi}_m \rangle$ represents the norm of $\boldsymbol{\Psi}_m$. The inner product constitutes a fundamental building block of standard (Hermitian) electrodynamics [Har61,Joa08] since it allows to project any scattered wave on the basis of modes, $\langle \boldsymbol{\Psi}_s | \boldsymbol{\Psi}_p \rangle = \alpha_p \langle \boldsymbol{\Psi}_p | \boldsymbol{\Psi}_p \rangle$. In particular, the essential orthonormalization of normal modes is rooted in all variational and perturbational methods. Physically, orthogonal and normalized field distributions enable a direct evaluation of the strength $\alpha_m$ with which a mode interact with an incident wave.

For conservative systems, the completeness and the definition of the inner product $\langle \boldsymbol{\Psi}_m | \boldsymbol{\Psi}_p \rangle$ directly follow from the properties of Hilbert spaces. However, non-Hermitian systems are described by non-self-adjoint operators whose eigenstates (the QNMs) do not form a Hilbert space. Therefore, the issues of completeness and orthonormalization with a suitable inner product are more complex.

The question of decomposing the response of a non-Hermitian system as a superposition of QNMs has a venerable history, which probably started in quantum mechanics [Gam28,Sie39,Zel61,Mor71, Mor73,Hoe79], see [Moi98,Moi11] for a recent review. Important developments have also been provided in the context of gravitational physics and general relativity [Pre71,Kok99,Kon11]. Note however that in this case the scattering potential has an infinite spatial extent, which causes specific difficulties not encountered in other fields of non-Hermitian physics. On the other hand, electromagnetism is a more favorable environment for QNM theories since the scattering potential (e.g., a permittivity change $\Delta\varepsilon$) has usually a compact support. The first studies in electromagnetism can be traced back to the seventies [Bau71,Bau76]. Let us emphasize that the vectorial nature of electromagnetic fields and the presence of a substrate or other complex environments raise specific issues that will be considered with care later on.

From the beginning, two fundamental question have received considerable attention. Under which conditions are QNM expansions complete? How can we evaluate the amplitude of each mode in a superposition of QNMs considering that it is challenging to properly define an inner product and an orthonormalization for modes whose field diverges as $|\mathbf{r}| \to \infty$? Two different frameworks have been developed in order to deal with these fundamental issues.

Since the nice properties of Hilbert space are lost, the derivation of orthonormalization is a long and winding road. Therefore, a first class of methods for deriving QNM expansions chose a framework radically different from the routine of Hermitian systems. Instead of relying on the definition of an inner product, this approach is based on the theory of functions of a complex variable, more particularly the residue theorem and one of its consequences, the Mittag-Leffler's theorem [Mor71,Mor73,Mul10,Doo14,Man17,Col18]. Under certain conditions, a representative function of the system response (often its Green tensor) is a meromorphic function that can be decomposed as a sum of its poles, which are the QNMs complex frequencies. It is interesting to note that this approach does not explicitly rely on the definition of an inner product. Consistently with the terminology adopted in [Lal18], we refer to this approach as the *residue-decomposition approach*.

On the contrary, the second approach for deriving QNM expansions closely follows the routine of Hermitian systems (the simple geometrical idea of projecting the system response), even if the vector space formed by the QNMs is not a Hilbert space. The regularization of the diverging fields is of particular importance in this approach in order to define a correct inner product that leads to orthogonality and normalization [Zel61,Leu94a,Moi98,Moi11,Sau13,Via14,Yan18,Tru20]. The guiding principle is to "map" the exponentially divergent QNMs into bounded square-integrable modes, which can be considered as new vectors of a sort of generalized Hilbert space. The important (and physically appealing) consequence of this mapping is that regularized QNMs share many similarities with normal modes of conservative systems. Consistently with the terminology adopted in [Lal18], we refer to this approach as the *orthogonality-decomposition approach.*

### 1.4 Motivation

QNM completeness, orthogonality, and normalization have been recently addressed, see the Sections 4 and 5 of the review article [Lal18] and the Sections 3 and 4 of the tutorial [Kri20]. However, the topics covered in these articles are much broader. On the contrary, the present review is focused on these issues and therefore delivers a much more comprehensive message. More specifically, the following points are developed in more details: (i) practical difficulties and limitations of the residue-decomposition approach (see the Section 2.3), (ii) necessary regularization of the QNM field that is at the heart of the orthogonality-decomposition approach (see the Section 3), and (iii) a comparison of different methods for normalizing QNMs (see the Section 4). The latter is accompanied by new mathematical demonstrations and numerical evidence. The global vision we intend to provide is, to our knowledge, unique in the field of electromagnetic QNMs. It may didactically lead newcomers in the field towards sound, concrete, and practical methods to calculate and use QNM expansions. It is especially important because some of these methods are already at the heart of free softwares [MAN,Dem20].

More specifically, the puzzling issue of normalizing diverging fields, which are not square integrable, has lastingly hindered the development of QNM formalisms for analyzing nanoscale light confinement. The issue is theoretically solved nowadays, see for instance [Lal18], at least in broad terms, owing to the intense efforts provided during the last decade. However, this has not been achieved in a progressive and linear way. As often in science, controversies have arisen along the way. For instance, the authors of [Kri15] discussed different normalization methods and concluded that they provide the same result. This conclusion was vividly debated in a series of articles, comment and reply [Sau15,Mul16a,Mul17,Kri17]. When preparing this review, we realized that the literature, including earlier milestone publications, is quite confusing, sometimes misleading, and not free of mistakes. As a result, more recent works take some erroneous conclusions for granted, while correct results seem to be already forgotten. We try to elucidate these points, while keeping the discussion accessible for a broad audience. We expect that the clarification will be helpful, not only for experts working in the field, but also for users who want to apply the QNM machinery to a variety of different geometries. We believe that it is instructive to clarify how ideas evolve during the process of building a theory to help newcomers and to promote further developments.

### 1.5 Structure

Besides the introduction, this review is divided up into five sections and is enriched by five appendices. Sections 2 and 3 address the construction of QNM expansions with the residue-decomposition approach and the orthogonality-decomposition approach. In general, the residue-decomposition approach works with the divergent QNMs of the open space (Section 2) while the orthogonality-decomposition approach requires regularized (i.e., non-divergent and square-integrable) QNMs (Section 3). Section 4 is devoted to the normalization issue. We first provide a historical perspective on the development of different normalization methods and then we compare the domain of validity of these methods. Section 5 summarizes and concludes the discussion.

Section 2 is devoted to the divergent QNMs of an open space. Section 2.1 first recalls the divergence issue that hinders the use of the classical inner product of conservative systems. Then, Section 2.2 presents the complexity of the spectrum (eigenvalues distribution in the complex frequency plane) of the open-space Maxwell operator. Section 2.3 describes the main lines of the construction of QNM expansions with the residue-decomposition approach. The conditions under which completeness is achieved are discussed. Finally, Section 2.4 further reviews the completeness issue.

In Section 3, we review fundamental theoretical tools to regularize the divergent field of QNMs. The regularization is at the heart of the orthogonality-decomposition approach since it allows the definition of an inner product that leads to orthogonality and normalization. Importantly, we bridge the latest developments with earlier related methods explored in other fields than electromagnetism, especially in quantum mechanics. Section 3.1 first provides a short overview from the non-Hermitian quantum mechanics literature, which has promoted similarity transformations to regularize QNMs. In electromagnetism, perfectly matched layers (PMLs) are extensively used since their birth in the 1990's as a mathematical and numerical tool to implement outgoing wave boundary conditions in scattering problems at real frequencies. Using PMLs for the eigenproblem at complex frequencies is a quite natural road to build regularized QNMs. Section 3.2 presents the main ingredients of PML regularization in the continuous infinite space, while Section 3.3 addresses the key issue of PML regularization in a finite discretized space. Section 3.4 discusses important properties of regularized QNMs: norm, (bi)orthogonality, and completeness with numerical modes. The orthogonality-decomposition approach can be extended to systems involving coupling between Maxwell's equations and other physical or chemical equations. Such problematics will probably occupy an increasingly important place in nanophotonics studies. The use of regularized QNMs for these multiphysics situations is addressed in Section 3.5.

Section 4 goes through the history of QNM normalization in optics, which can be separated in two main periods. In the 1990's, P. T. Leung, K. Young, and coworkers at the Chinese University of Hong Kong played an important role in the development of QNM theory for optical cavities. They studied scalar fields in 1D cavities as well as vector fields in 3D systems with a spherical symmetry. Then, the 2010's witnessed a second active period where different groups addressed the issue of QNM normalization, for arbitrary 3D geometries for which no analytical solution exists.

In view of this historical perspective, one can underline that a few general normalization methods exist since typically ten years to normalize the QNMs of any general structure, potentially made of dispersive, anisotropic, magnetic materials and surrounded by non-uniform media, as is the case with resonators on substrates for instance. The domain of validity of these different normalization methods in explored in detail. In addition, we correct imprecisions or inconsistencies of the recent literature. In particular, we recall that the normalization method inherited from the works of the Hong Kong group is only valid for 1D systems and provide erroneous results for any other case, even 3D systems with a spherical symmetry [Mul16a,Mul17].

Finally, let us briefly describe the content of the Appendices. Appendix A recalls the Lorentz reciprocity theorem, sometimes called divergence theorem, because it plays a central role in the review. We provide in Appendix B a proof of the invariance of the QNM norm calculated with PMLs. The review mostly considers the case of reciprocal materials. However, we show in Appendix C that the PML regularization allows to treat non-reciprocal materials as well. Appendix D addresses the numerical implementation of the normalization.

## 2.  The divergent QNMs of the open space

In optics and more generally in electromagnetism, the QNMs of any open cavity are formally defined as the time-harmonic solutions to the source-free Maxwell's equations with outgoing waves boundary conditions [Lal18]

$$\begin{aligned}\nabla \times \widetilde{\mathbf{E}} &= i\widetilde{\omega}\boldsymbol{\mu}(\mathbf{r},\widetilde{\omega})\widetilde{\mathbf{H}} \\ \nabla \times \widetilde{\mathbf{H}} &= -i\widetilde{\omega}\boldsymbol{\varepsilon}(\mathbf{r},\widetilde{\omega})\widetilde{\mathbf{E}} \\ + \text{ outgoing waves conditions at infinity} &\end{aligned} \quad . \tag{1}$$

We use throughout the article the $\exp(-i\omega t)$ convention and denote with a tilde QNM complex-valued frequencies, $\widetilde{\omega} = \Omega - i\Gamma/2$, where $\Gamma$ stands for the decay rate, i.e., the inverse of the mode lifetime $\tau = 1/\Gamma$. The factor 2 accounts for the difference between amplitude and energy decays.

In Eq. (1), $\boldsymbol{\varepsilon}(\mathbf{r},\omega)$ and $\boldsymbol{\mu}(\mathbf{r},\omega)$ are the permittivity and permeability distributions of the system under study. They are in general frequency-dependent and complex-valued tensors characterizing dispersive, absorptive, and anisotropic materials. We assume them to be equal to their transpose (reciprocal materials) for simplicity reasons. The specific case of non-reciprocal material is studied in Appendix C. We do not consider situations involving coupling between Maxwell's equations and other physical or chemical equations, e.g., optomechanics or thermoplasmonics, but the following developments remain valid if the joint operator combining all differential equations under consideration admits an analytical continuation at complex-valued frequencies. Incidentally, note that when Drude or Lorentz permittivities are considered, multiphysics is already enrolled in Eq. 1 (electromagnetism and charge motion).

## 2.1 Divergence of the QNM field

The non-Hermitian character of QNMs has a thorny consequence: the exponential growth of the QNM field away from the resonator. It is easy to understand this divergence by considering a resonator embedded in a uniform medium with a refractive index $n$. Since the QNMs satisfy the outgoing-wave condition, their fields take for large $r$'s the asymptotic simple form of a progressive wave with a spherical wavefront, $\mathbf{A}(\theta,\varphi)r^{-1}\exp[i\widetilde{\omega}(-t+nr/c)]$. Since the QNM fields have to exponentially decay in time, $\text{Im}(\widetilde{\omega}) < 0$ and the fields exponentially diverge in space as $r \to \infty$ [Chr19].

Despite what is occasionally read in the literature, the exponential growth does not raise any physical inconsistency. It follows from causality: QNMs are exponentially damped in time. Thus, they diverge as $t \to -\infty$, and since the QNM field radiated at $t = -\infty$ is perceived at $r \to \infty$ at a finite $t$, the QNM field diverges in space. The spatial divergence for $r \to \infty$ is not more unphysical than the temporal divergence for $t = -\infty$. In practice, the signal does not start at $t = -\infty$, the field is actually observed over finite spatial intervals, and the divergence is not met.

On another note, the divergence of the electromagnetic field represents a difficulty for the development of a non-Hermitian electrodynamic formalism. The concept of inner product constitutes a fundamental building block of standard (Hermitian) electrodynamics [Har61,Joa08]. In particular, it leads to the essential orthonormalization of normal modes that is rooted in all variational and perturbational methods based on the field expansion over a set of eigenstates. Indeed, the inner "wavefunctions" product of Hermitian electrodynamics $\int \widetilde{\mathbf{E}}_m \cdot \widetilde{\mathbf{E}}_n^* \, d^3r$ cannot be used in the non-Hermitian context. Furthermore, note that the unconjugated form $\int \widetilde{\mathbf{E}}_m \cdot \widetilde{\mathbf{E}}_n \, d^3r$ often used for the orthonormality of guided (bounded) modes in absorbing waveguides [Sny83] cannot be used as such either, since the QNMs diverge as $r \to \infty$.

## 2.2 The spectrum of Maxwell's operators

Figure 1(a) shows the spectrum (eigenvalues distribution in the complex frequency plane) of the open-space operator of Eq. (1). The spectrum comprises a discrete set of QNMs represented with red dots. They are often termed natural frequencies since they are frequencies for which the object can have a response even with no excitation by an incident wave. If the object is excited at a natural frequency, then its response has an infinite amplitude at that complex frequency. They are thus poles of the scattering operator. Owing to the Hermitian symmetry of real Fourier transforms, $\boldsymbol{\varepsilon}^*(\mathbf{r},\omega) = \boldsymbol{\varepsilon}(\mathbf{r},-\omega^*)$ and $\boldsymbol{\mu}^*(\mathbf{r},\omega) = \boldsymbol{\mu}(\mathbf{r},-\omega^*)$. Thus, if $(\widetilde{\mathbf{E}}_m, \widetilde{\mathbf{H}}_m, \widetilde{\omega}_m)$ is a source-free solution of Maxwell's equations, $(\widetilde{\mathbf{E}}_m^*, \widetilde{\mathbf{H}}_m^*, -\widetilde{\omega}_m^*)$ is another solution for the same permittivity and permeability distribution, see the Section 2.3 in [Lal18]. This explains why every eigenfrequency $\widetilde{\omega}$ in the lower right quadrant of the complex plane is paired with another twin frequency $-\widetilde{\omega}^*$ in the lower-left quadrant, except those with a null real part.

The spectrum may feature other singularities, such as accumulation points arising from either poles of the $\varepsilon(\omega)$ or $\mu(\omega)$ (material resonance) [Bru16,Mul16b,Man17,Yan18,Tru20,Sau21], corner modes [Dem20], or plasmonic resonances [Bru16,Yan18,Tru20]. The latter appear in the case of metallic particles surrounded by a dielectric material and corresponds to complex frequencies for which $\varepsilon(\omega) = -\varepsilon_b$, with $\varepsilon_b$ the permittivity of the dielectric medium. This kind of accumulation point is related to quenching, a fundamental and detrimental process for a molecule that emits light in the vicinity of a metallic material, see Fig. 3 in [Yan18].

The spectrum may also feature branch cuts when the resonator is not surrounded by a uniform medium, a typical example being a resonator laid on a substrate. Different outgoing wave conditions should be satisfied in two different half spaces, the substrate and superstrate. The difficulty further increases for resonators surrounded by complex environments that potentially support guided waves. Typical examples are ring resonators, photonic-crystal cavities, or plasmonic nanoantennas coupled to a waveguide. The last case is analyzed in Figs. 3 and 4.

Finally, let us also mention spectral singularities, known as exceptional points, for which two or more QNMs coalesce for specific values of an opto-geometrical parameter. Such degeneracies have recently raised considerable attention in photonics, given that optical gain and loss can be used as nonconservative ingredients to create the right conditions for the apparition of exceptional points, which in turn result in new optical properties [Par20].

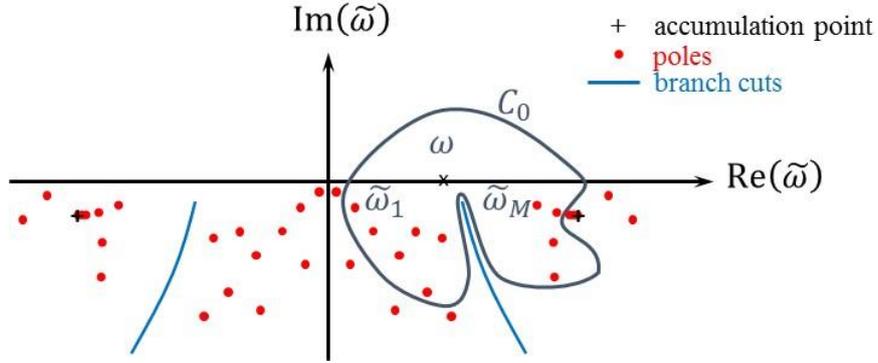

**Fig. 1.** Complex frequency plane for a resonant open system. The problem is defined by the Maxwellian operator of Eq. (1) in an unbounded space with real spatial coordinates. The spectrum features accumulation points (black plusses), QNMs (red dots), branch cuts (blue solid curve). $C_0$ is a closed and finite counterclockwise oriented curve. It is chosen to encircle the point $z = \omega$ (often on the real axis) and a finite set of poles, $\widetilde{\omega}_m, m = 1, \ldots M$, but to avoid branch cuts.

## 2.3 QNM expansion and completeness

To interpret experimental results involving resonant structures, physicists like to consider how the resonance explicitly impacts the system, reconstructing the resonator response (at least in a compact subspace of $\mathbb{R}^3$) with a QNM expansion of the scattered field

$$\begin{bmatrix} \mathbf{E}_S(\mathbf{r},\omega) \\ \mathbf{H}_S(\mathbf{r},\omega) \end{bmatrix} = \sum_m \alpha_m(\omega) \begin{bmatrix} \widetilde{\mathbf{E}}_m(\mathbf{r}) \\ \widetilde{\mathbf{H}}_m(\mathbf{r}) \end{bmatrix}, \qquad (2)$$

in the frequency domain. The intrinsic force of modal expansions is that they provide key clues towards the understanding of the physics of the resonator. In Eq. (2), the $\alpha_m$'s are the modal excitation coefficients, which describe how much the QNMs are excited when the resonator is driven by a monochromatic incident beam at frequency $\omega$. A similar expression exists in the time domain when the resonator is driven by a pulse [Lal18]. Throughout the article, we refer to Eq. (2) as a QNM expansion, and if the scattered field can be exactly reconstructed with QNMs, we will say that the QNM basis forms a complete basis.

There is a large literature on QNM expansions, and their completeness is considered as a crucial question that has commanded considerable attention in foundational electromagnetic studies [Bau71,Bau76,Mor71,Mor73]. It is interesting to summarize the main results, and for that purpose, we follow the residue-decomposition approach based on the pole expansion of meromorphic functions (Mittag-Leffler theorem) [Arf05]. Equation (2) links a function of a real variable $\omega$ to its complex poles, and in complex analysis, this is achieved by evaluating line integrals of analytic functions over closed curves. Let us consider the integral

$$\frac{1}{2\pi i} \int_{C_0} \frac{\mathbf{E}_S(\mathbf{r},z)}{z-\omega} dz, \qquad (3)$$

which is taken along a closed and finite counterclockwise oriented curve $C_0$, see Fig. 1(a). The contour encircles the point $z = \omega$, which is a pole of the function $\frac{\mathbf{E}_S(\mathbf{r},z)}{z-\omega}$, and a finite set of poles, $\widetilde{\omega}_m, m = 1, \ldots M$, of the function $\mathbf{E}_S(\mathbf{r}, z)$. Assuming that the poles are all of order 1 and that the function $\frac{\mathbf{E}_S(\mathbf{r},z)}{z-\omega}$

is holomorphic inside a domain $U\backslash\{\omega, \widetilde{\omega}_1 \ldots \widetilde{\omega}_M\}$ containing $C_0$ for a position **r**, the residue theorem provides

$$\int_{C_0} \frac{\mathbf{E_S}(\mathbf{r},z)}{z-\omega} dz = 2\pi i \sum_{k=1\ldots M} \text{Res}\left(\frac{\mathbf{E_S}(\mathbf{r},z)}{z-\omega}, \widetilde{\omega}_k\right) + 2\pi i \mathbf{E_S}(\mathbf{r},\omega), \tag{4}$$

which is conveniently rewritten

$$\mathbf{E_S}(\mathbf{r},\omega) = \frac{1}{2i\pi}\int_{C_0} \frac{\mathbf{E_S}(\mathbf{r},z)}{z-\omega} dz - \sum_{k=1\ldots M} \text{Res}\left(\frac{\mathbf{E_S}(\mathbf{r},z)}{z-\omega}, \widetilde{\omega}_k\right), \tag{5}$$

where $\text{Res}(f, \widetilde{\omega}_k)$ denotes the residue of the function $f$ at the pole $\widetilde{\omega}_k$. Note that $\text{Res}\left(\frac{\mathbf{E_S}(\mathbf{r},z)}{z-\omega}, \widetilde{\omega}_k\right) = \text{Res}(\mathbf{E_S}(\mathbf{r},z), \widetilde{\omega}_k)/(\widetilde{\omega}_k - \omega)$. Equation (5) is valid if the singularities inside $C_0$ are the poles only, implying that $C_0$ rounds all branch cuts to avoid crossing them.

To analyze the question of completeness, one enlarges the contour $C_0$ so that all the QNM poles are encircled. The line integral of the first term on the right-hand side of Eq. (5) is now computed over a circle denoted $C_\infty$ with a center at $\omega = 0$ and an infinitely large radius [Mor71]. If the integral over the circle $C_\infty$ is zero, i.e., if the function $\frac{\mathbf{E_S}(\mathbf{r},z)}{z-\omega}$ vanishes sufficiently fast as $|z| \to \infty$ and if there is no branch cuts, we then end up with the desired expansion for which the response is set as a sum of independent pole contributions

$$\mathbf{E_S}(\mathbf{r},\omega) = \sum_{k=1\ldots\infty} \frac{\text{Res}(\mathbf{E_S}(\mathbf{r},z), \widetilde{\omega}_k)}{\omega - \widetilde{\omega}_k}. \tag{6}$$

and the basis of the QNMs is said to be complete.

In general, the integral over $C_\infty$ is zero as long as the parameter **r** belongs to a certain compact region and the QNM basis is then complete only for positions **r** inside this region. The region for which the QNM expansion is complete has been largely studied for simple problems for which the scattered field or the Green tensor is known with a closed form expression [Bau71,Bau76,Mor71,Mor73]. 1D Fabry-Perot resonators (i.e., diode junctions in quantum mechanics) or 3D spherical Mie-resonators have been primarily analyzed. For 1D Fabry Perot resonators, typically a 1D slab of permittivity $\varepsilon(x) = \varepsilon_0$ for $|x| < L/2$ in a uniform background, $\varepsilon(x) = \varepsilon_1$ for $|x| > L/2$, the QNM basis is complete inside the slab in the interval $]-L/2; L/2[$ [Leu94a]. For spherical resonators defined by $\varepsilon(r) = \varepsilon_0$ for $|r| < R$ and $\varepsilon(r) = \varepsilon_1$ for $|r| > R$, completeness is achieved inside the sphere of radius $R$ [Leu96]. These results are quite intuitive; at least, it is obvious to see that the field outside the cavity behaves as $\exp[i\widetilde{\omega}nr/c]$, and thus exponentially diverges as $z \to \infty - i\infty$ on the lower half circle of $C_\infty$. From these initial studies, it is generally admitted nowadays that completeness is achieved inside the minimal sphere that surrounds the resonator, outside which the permittivity is uniform. However, a rigorous proof of the vanishing of the integral over $C_\infty$ inside the resonators with arbitrary geometries is still waiting to be done. More details can be found see the Section 4 and Appendix C in [Kri20].

Disappointingly, the line integral over $C_\infty$ is not equal to zero in general and Eq. (6) is rarely met in practice. Indeed, most systems [note1] involve a boundary between two media, defining a substrate and a superstrate, e.g., a metal particle deposited on a substrate or a photonic-crystal cavity in a semiconductor membrane. The outgoing-wave condition has to be satisfied into two different half spaces and two branch cut singularities appear. The circle $C_\infty$ should then be distorted to avoid crossing the branches, as illustrated in Fig. 1, and the contour integral is not null. One may argue that the contribution of the branch cut may also be computed numerically by integration in the complex plane along the distorted contour $C_\infty$, but in practice this is not easy to implement and it has only been achieved for very simple geometries where the branch-cut contribution can be analytically obtained or when all the poles can be computed [Doo13].

We note that a precise knowledge of the spatial region for which the QNM basis is complete is not critical nowadays, since this knowledge is useful only for 1D or 3D resonators [note2] in uniform backgrounds, which are more of academic than practical interest [note1]. Further note that in recent works for 1D Fabry-Perot resonators [Abd18] and spheres [Col18], completeness is ensured outside the resonator, see also [Jar21] for the case of gravitational waves. These interesting results are obtained through a regularization based on a coordinate transformation of both the spatial and temporal coordinates. The latter has not received a generalization for higher dimensions so far, except for spheres that represent a direct generalization of 1D cases.

2.4 Conclusion

The first line in Table 1 summarizes the discussion on the completeness for the open space. The latter is guaranteed inside the resonator for geometries with uniform backgrounds only. The presence of branch cuts is problematic. When the function is not known analytically, it is not easy to compute the branch-cut contributions along the deformed contours. Note that the contribution depends on the scattered field itself. As a result, even if the geometry is fixed, it changes as the driving source is modified. We are not aware of earlier works that compute the branch-cut contribution, except for the specific case of 2D resonators in free space (without substrate) for which the branch cut arises from the singularity of the Green tensor. This case is conveniently much simpler than the case of branch cuts arising from the presence of a substrate, or a semiconductor slab as will be analyzed in Fig. 3.

Table 1. Completeness of the QNMs in unregularized (open) and regularized spaces.

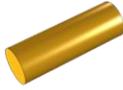

|  | Resonator in free space | Resonator on a substrate | Resonator in a PhC membrane |
|---|---|---|---|
| Open space (Eq. 1) | complete inside (i) | incomplete (ii) | incomplete (ii) |
| Bounded regularized space (Eq. 6) | complete everywhere (iii) | complete everywhere (iii) | ? (iv) |

(i) The proof of the completeness (inside the resonator only) has been notably obtained for 1D rods and 3D spherical resonators [Bau76,Mor71,Mor73,Leu94a].
(ii) Completeness is not achieved, neither outside nor inside the resonator.
(iii) Completeness is achieved by combining QNMs and numerical modes [Via14,Zol18,Yan18,Gra19,Gra20].
(iv) QNM normalization has been rigorously achieved [Fag17,Kri14a], but there is no study on the completeness owing to the lack of efficient QNM solver for 'periodic substrate', which prevents implementing regularization in a systematic way.

So, what to do in practice? The literature, as a whole, proposes to use a restricted finite subset of QNMs, and to complement this subset by another contribution that might be called a non-resonant contribution. The expansion may then be written as

$$\begin{bmatrix} \mathbf{E}_S(\mathbf{r},\omega) \\ \mathbf{H}_S(\mathbf{r},\omega) \end{bmatrix} = \sum_{m=1\ldots M} \alpha_m(\omega) \begin{bmatrix} \tilde{\mathbf{E}}_m(\mathbf{r}) \\ \tilde{\mathbf{H}}_m(\mathbf{r}) \end{bmatrix} + \begin{bmatrix} \tilde{\mathbf{E}}_{nr}(\mathbf{r}) \\ \tilde{\mathbf{H}}_{nr}(\mathbf{r}) \end{bmatrix}, \qquad (7)$$

where $M$ relevant QNMs are retained in the expansion. By 'relevant', we mean that they take a dominant impact on the physics. Let us emphasize that, to be useful, the equality in Eq. (7) should be achieved for any position $\mathbf{r}$.

Several approaches have been experimented to evaluate the non-resonant contribution. In the Riesz projection method [Zsc18,Bet21], a finite contour is considered, similar to $C_0$ in Fig. 1. The non-resonant contribution is accessed by computing $\mathbf{E}(\mathbf{r},z)$ for complex frequencies $z$ along the contour $C_0$, which requires solving Maxwell's equations rigorously typically one hundred times [Zsc18]. High accuracy is achieved in general; note however that the case with branch cuts has not been tested so far [Bur21]. Alternatively, the non-resonant contribution may also be computed by directly solving

Maxwell's equations at a few real frequencies, and since it is expected to gently vary with the frequency, it can be easily interpolated with the help of the real-frequency data [Wei18,Wu21b]. High accuracy has been demonstrated with this hybrid method for complex geometries, a dielectric grating [Wei18] and a silicon nanodisk on a metal thin layer [Wu21b].

The previous approaches have an obvious practical interest, as they may enable fast, intuitive and accurate methods for analysing the responses of resonators. However, they lack from the reassuring consistency provided by completeness. For instance, when QNMs are used to enable a coupled mode theory for the coupling of complex resonant scatterers with power-carrying channels or for the hybridization of several resonators, we cannot resort to real-frequency data, and would simply enjoy benefiting from a convergence by increasing the number of QNMs in the expansion. The same could be said also for the resonant-state expansion [Mul10,Doo14,Mul16b,Mul18], a rigorous high-order perturbation method for computing the quasinormal modes of any resonators by using a complete set of QNMs of a simpler system as a basis through a matrix diagonalization.

In the last few years, it has become clear that alternative formalisms relying on expansions that are complete inside and outside the resonators may be developed [Via14,Yan18,Gra20]. In this case, completeness is achieved by QNM regularization and by complementing an infinite set of QNMs by another infinite set of numerical modes (often called PML modes [note3]). The expansion takes the following form

$$\begin{bmatrix} \mathbf{E}_S(\mathbf{r},\omega) \\ \mathbf{H}_S(\mathbf{r},\omega) \end{bmatrix} = \sum_{m=1\ldots\infty} \alpha_m(\omega) \begin{bmatrix} \widetilde{\mathbf{E}}_m(\mathbf{r}) \\ \widetilde{\mathbf{H}}_m(\mathbf{r}) \end{bmatrix} + \sum_{p=1\ldots\infty} \alpha_p^{num}(\omega) \begin{bmatrix} \widetilde{\mathbf{E}}_p^{num}(\mathbf{r}) \\ \widetilde{\mathbf{H}}_p^{num}(\mathbf{r}) \end{bmatrix}. \qquad (8)$$

In Eq. (8), the numerical modes $\left[\widetilde{\mathbf{E}}_p^{num}(\mathbf{r}), \widetilde{\mathbf{H}}_p^{num}(\mathbf{r})\right]^T$ are denoted just like the actual QNM and their excitation coefficients also (we simply add a superscript '$num$' for the sake of differentiation). The reason is that the numerical modes are computed exactly as the QNMs and that their excitation coefficients have a closed form expression, which is exactly the same as that of the QNMs.

Importantly, we note that, since the completeness of the expansion is guaranteed inside or outside the resonators, the formalism is particularly suited to various coupled mode theories, including coupling between scattering bodies and waveguide channels [Zha20], hybridization of nanoresonators [Cog20,Tao20] and rigorous high-order perturbation theory for large shape deformation [Yan20].

The present review is not concerned by formulas for the $\alpha$'s [Lal18,Gras20]. Let us mention that there is a single valid expression of the $\alpha$'s for resonators made of non-dispersive materials for a given inner product [note4], since the latter allows us to project $[\mathbf{E}_S, \mathbf{H}_S]^T$ onto the basis vectors $\left[\widetilde{\mathbf{E}}_m, \widetilde{\mathbf{H}}_m\right]^T$ [Sau13,Lal18]. The case of dispersive resonators will be briefly mentioned see the Section 3.4.3.

The expansion in Eq. (8) fundamentally relies on the regularization of QNM. We now examine its implication.

## 3. Regularized QNMs

Let us start by recalling that the concept of the inner product constitutes a fundamental building block of standard (Hermitian) electrodynamics or quantum mechanics in classical textbooks on conservative systems, see Chapter 10 in [Arf05]. With inner products, one may conveniently normalize the vectors, construct an orthogonal basis and perform projections onto one vector of the basis.

Indeed, one anticipates that, if we may define a sort of inner product for non-Hermitian systems, one may then accommodate the tools of conventional Hermitian electrodynamics and considerably simplify the treatment of non-Hermitian system. This is exactly the approach that is highlighted in this section, which will end up with the definition of regularized QNMs and numerical modes, which all obey the same orthogonality product and together form a complete basis of a regularized Hilbert space with square-integrable vectors.

The concept of QNM regularization has been initially developed in a series of theoretical studies in quantum mechanics, which will be reviewed in the Section 3.1. We will then focus on QNMs that are regularized with perfectly matched layers, first with an infinite thickness see the Section 3.2, and then with a finite one in the Section 3.3. Finite thicknesses correspond to the current practice in computational electrodynamics. This will give us the opportunity to well differentiate the spectra of continuous operators of infinite space with those of discrete operators of discretized and finite space. We will then conclude the Section by commenting on the completeness of the basis formed by regularized QNMs and numerical modes, and the different tests that are convincingly supporting the completeness.

3.1 An overview from the quantum mechanics literature

From the beginning of theoretical study on the spectral decomposition of scattered waves into a superposition of QNMs, a fundamental question has drawn considerable attention: How may we properly "map" the divergent QNMs of the open space onto a Hilbert space and force the QNMs to be square-integrable. The general recipe, widely developed in quantum mechanics and gravitational physics, consists in carrying out a suitable mathematical mapping that converts the original exponentially divergent QNMs into bounded square-integrable QNMs, which can be considered as new vectors of a new Hilbert space that shares many similarities with the normal modes of conservative systems.

In this overview section, we adopt the presentation of Chapter 5 of the monograph in [Moi11] which emphasizes mappings with similarity transformations. For the sake of easy read by an interested reader, we symbolically recast the continuous Maxwellian operator of Eq. (1) into the classical Hamiltonian format $\mathbf{H}\widetilde{\boldsymbol{\Psi}} = \widetilde{E}\widetilde{\boldsymbol{\Psi}}$ of quantum mechanics, with $\widetilde{E} \equiv \hbar\widetilde{\omega}$ the complex energy and $\widetilde{\boldsymbol{\Psi}} \equiv [\widetilde{\mathbf{E}}; \widetilde{\mathbf{H}}]$ the wavefunction. The similarity transformation

$$\widetilde{\boldsymbol{\Psi}}_r = \mathbf{U}\widetilde{\boldsymbol{\Psi}} \tag{9}$$

maps the divergent QNMs $\widetilde{\boldsymbol{\Psi}}$ ($|\widetilde{\boldsymbol{\Psi}}| \to \infty$ for $r \to \infty$) into new 'regularized' QNMs $\widetilde{\boldsymbol{\Psi}}_r$ ($|\widetilde{\boldsymbol{\Psi}}_r| \to 0$ for $r \to \infty$) and consists in solving a new spectral problem

$$\mathbf{H}_r\widetilde{\boldsymbol{\Psi}}_r = \widetilde{E}\widetilde{\boldsymbol{\Psi}}_r \tag{10a}$$

$$\mathbf{H}_r = \mathbf{U}\mathbf{H}\mathbf{U}^{-1}, \tag{10b}$$

with identical eigenvalues $\widetilde{E}$. Note the subscript '$r$' (for 'regularized') used to differentiate the new spectral problem from the initial unscaled one.

The key point is how to find such a linear mapping. The literature on non-Hermitian quantum mechanics contains various approaches in which such a transformation is introduced. We will not review them and refer the interested readers to Chapter 5 in [Moi11] for a comprehensive review, starting with the Gaussian regularization introduced by Zel'dovich in the 1960's [Zel61,Baz69] and recently revisited in [Sto21] for Maxwellian operators. Most of the approaches are in fact rather mutually equivalent. They are often applicable for simple analytical potentials (scalar waves in 1D) and are inconveniently transferable to contemporary electromagnetic resonators encountered in nanophotonics.

The motivation for applying the similarity transformation is obvious. Under this transformation, the resonance wavefunctions initially exponentially growing up in space become 'regular' functions that are square-integrable. One immediately anticipates that normalization and orthogonality may then be defined in the new Hilbert space of regularized functions.

In electromagnetism, the royal road to build regularized QNMs is that provided by perfectly matched layers (PMLs). Owing to the importance of scaled spaces for deriving the orthonormality relation, the PML machinery will be carefully documented in the next subsection. We simply note, at this stage, that a PML does not provide a one-to-one mapping between the divergent and regularized QNMs, as Eq. (2) simplistically suggests.

As an aside, let us note that another regularization approach has been recently introduced in the literature on QNMs [Ge14,Dez18] and further utilized in a series of applications, including the important case of QNM quantization [Fra19,Fra20]. The regularization relies on a dual scheme, with the embodiment of a QNM field evaluated at the real-valued resonance frequency of the QNM outside the resonator while keeping the actual QNM field evaluated at the complex-valued frequency inside the resonator. The divergence is therefore removed by construction. The heuristic is simple and removes all divergence issues with a wave of a magic wand. For dispersive media, the regularized QNM no longer satisfies the interface conditions for tangent electromagnetic fields at the resonator boundary. More fundamentally, the heuristic does not preserve the machinery of analytic continuation, which is central in QNM theory, and fundamental notions such as completeness and orthogonality are not preserved. In our opinion, it is a quick fix that deserves more in-depth considerations.

## 3.2 PML-regularization in infinite space

Usually, PMLs are used for damping oscillating waves corresponding to scattered fields at real frequencies. Here we review how they can be used for the regularization and normalization of QNMs. A convenient and efficient method to divert the divergent resonance QNM into the physical domain of square-integrable wavefunctions is to rotate the coordinate along which the divergence occurs into the complex plane by an angle $\theta$.

To guide the analysis, Fig. 2(a) shows a 2D resonator on a complicated substrate. On purpose, the "substrate" is composed of two semi-infinite uniform half-planes with different permittivities. Importantly, we note that for $x$ large enough, i.e. for $|x|$ larger than some positive coordinate $x_0$, the permittivity is invariant in $x$; it depends only on $y$. Conversely, the same remark can be made for the $y$ coordinate. The key fact about these geometries that already encompass a great variety of practical cases is that the solution $\widetilde{\boldsymbol{\Psi}} = [\widetilde{\mathbf{E}}; \widetilde{\mathbf{H}}]$ of Maxwell's equations is an analytic function of the spatial coordinate $\mathbf{r}$ (see for instance [Hug05]), which can be freely analytically continued and rotated at complex values of $\mathbf{r}$ with PMLs [Che94,Ber07].

The figure illustrates a possible PML implementation composed of three different regions. In the center part, no coordinate transform is applied; the physical space is unmapped. In the regions along $x$ or $y$ axis, only a single direction is mapped; finally, at the corners, both the $x$ and $y$ coordinates are mapped, independently. This complex rotation that keeps the coordinates unscaled in the resonator region where the permittivity is complicated is referred to as the exterior scaling transformation in quantum mechanics studies; it has been introduced in the late 70's, see the Section 5.3 in [Moi11].

The following analysis can be conducted for the 2D geometry of Fig. 2(a) by detailing all the PML regions [Hug05], but for simplicity reasons, we will adopt a simplified analysis, just highlighting the main idea. First, we note that one moves from any rectangular region to a neighbor one in Fig. 2(a), just by applying or removing a single coordinate transform in a single direction, either $x$ or $y$. Therefore, for the following analysis, we consider that the mapping is performed along a single coordinate, denoted $x$, hereafter. Furthermore, we consider a mapping for $x \to +\infty$ only, the other Cartesian coordinates and $x < 0$ being treated similarly. By hypothesis (see Fig. 2(a)), the permittivity and permeability are both independent of $x$ for $x > x_0$. Thus, the QNM in infinite space may be decomposed as a superposition of waves of the form $\boldsymbol{\Psi}_k(y,z)\exp(i\tilde{k}_x x - i\widetilde{\omega}t)$ (modal expansion with radiation and guided waves) [Hug05] or simply planes waves if the half space $x > x_0$ is uniform for any $y, z$ (Helmholtz–Weyl decomposition).

Let us first consider the field $\boldsymbol{\Psi}(x,y,z)$ scattered by the resonator under illumination by a driving field at real frequency $\omega$. Since $\boldsymbol{\Psi}(x,y,z)$ satisfies the outgoing wave condition, $\boldsymbol{\Psi}(x,y,z)$ gently decays as $1/r$ and oscillates at large distances $r$ from the resonator. Thus, $\boldsymbol{\Psi}$ admits an analytical continuation $\boldsymbol{\Psi}(\tilde{X},y,z)$ in the complex domain defined by $\text{Re}(\tilde{X}) > x_0$ ($x_0$ being the inner boundary of the scaled exterior) and $\text{Im}(\tilde{X}) > 0$, with $\boldsymbol{\Psi} \to 0$ as $\tilde{X} \to \infty + i\infty$. The Maxwell's equations, initially restricted to the real axis $x = \text{Re}(\tilde{X})$, can be solved in this infinite complex space along an appropriate trajectory with, for instance, a linearly growing imaginary part, see Fig. 2. This is the essence of the PML concept [Che94,Ber07]. We adopt the presentation and notation of [Hug05] hereafter.

The key point is that, owing to the choice $\text{Im}(\tilde{X}) > 0$, only the outgoing waves admits an analytical continuation, since all propagative or evanescent waves of the form $\boldsymbol{\Psi}_k(y,z)\exp(-ikx - i\omega t)$ exponentially diverge as $\exp(-ik\tilde{X} - i\omega t) \to \infty$ for $\tilde{X} \to \infty + i\infty$. Thus, by construction, PMLs automatically implement the outgoing wave condition. Additionally, since the analytical continuation guarantees that no reflection occurs at $\text{Re}(\tilde{X}) = x_0$, PMLs (with an infinite thickness as we investigate in this Section) *perfectly* satisfy the outgoing wave conditions by damping the field without reflection, at least for many geometries of wide interest [note5]. They are often called reflection-less absorbing materials and this is why they have revolutionized boundary conditions in electromagnetism and other wave equations. The next Section considers finite-thickness effect.

The integration along a trajectory with a complex coordinate is inconvenient. Thus, we conveniently *change variables* and map the complex trajectory into a new one depending only on a real variable, $x' = \text{Re}(\tilde{X})$ [note6]. The mapping requires to know the relationship between $\text{Im}(\tilde{X})$ and $\text{Re}(\tilde{X})$, i.e. the function $\tilde{X} = F(x')$. Owing to the equivalence between material parameters and spatial coordinate transforms [Nic94,Leo06], the mapped Maxwell equations can be either written with transformed operators by replacing $\frac{\partial(\cdot)}{\partial \tilde{X}}$ by $\frac{\partial x}{\partial \tilde{X}} \frac{\partial(\cdot)}{\partial x}$ in Eq. (1) or by using transformed permittivity and permeability tensors, or by a combination of both approaches [Hug05].

Figure 2(b) shows the simplest possible mapping corresponding to $\tilde{X} = x'$ for $x' < x_0$ and $\tilde{X} = x' + i \tan(\theta)(x' - x_0)$ for $x' > x_0$ ($x'$ and $x$ coincides in this simple case). This transform defines a rotational angle $\theta$ of the coordinates in the complex plane, $\tan(\theta) = \text{Im}(\tilde{X})/(x' - x_0)$. Other convenient mappings involve complex coordinate stretching, equivalent to an additional change of the length units, or dispersive coordinate transform [Ber07]. They are unessential for our discussion.

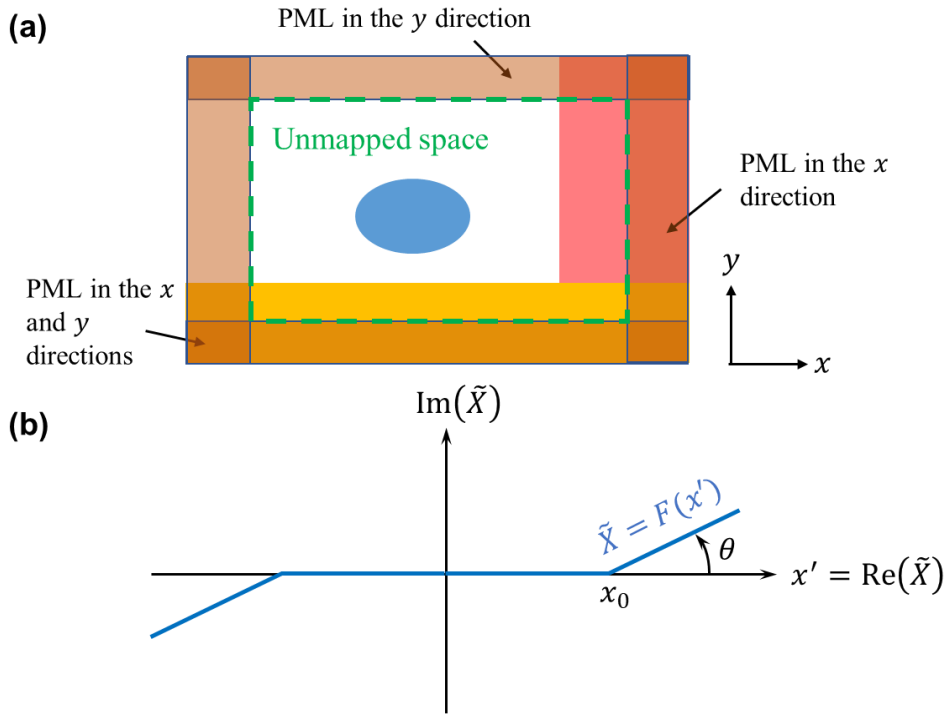

**Fig. 2.** Complex mappings with PMLs. **(a)** Illustration of a PML implementation in 2D for a resonator with a complicated "substrate" composed of two semi-infinite uniform orthogonal half planes with different permittivities. Importantly, we note that the PMLs are implemented by complex mappings that can all be factorized as the product of a function of $x$ or $y$ only. **(b)** Example of an appropriate trajectory in the complex coordinate plane that is classically used in the PML machinery. The key point is that the trajectory encompasses a linearly growing imaginary part for $\tilde{X} \to \infty + i\infty$. Note that no mapping is applied for $|x| < x_0$, implying that the regularized QNMs defined by Eq. (11) and the initial diverging ones defined by Eq. (1) coincide in this bounded region of space. (b) is adapted from [Leu94a].

Let us come back to the QNM solution of Eq. (1) with a complex frequency $\widetilde{\omega}$. Just like for real frequencies, the PML machinery can be applied. Using the transformed permittivity and permeability tensors, we may symbolically rewrite the regularized spectral problem of Eq. (1)

$$\nabla \times \widetilde{\mathbf{E}}(\mathbf{r}') = i\widetilde{\omega}\boldsymbol{\mu}'(\mathbf{r}',\widetilde{\omega})\widetilde{\mathbf{H}}(\mathbf{r}')$$
$$\nabla \times \widetilde{\mathbf{H}}(\mathbf{r}') = -i\widetilde{\omega}\boldsymbol{\varepsilon}'(\mathbf{r}',\widetilde{\omega})\widetilde{\mathbf{E}}(\mathbf{r}'), \qquad (11)$$

where '$\nabla$' simply denotes the curl operator acting on the mapped coordinate $\mathbf{r}' = (x',y',z')$. The prime indicates that the tensors have been changed to account for the complex coordinate transform.

There is however a minor, albeit essential, difference with the usual real-frequency case: QNM fields diverge as $\exp(i\tilde{k}x - i\widetilde{\omega}t) \equiv \exp(-\text{Im}(\tilde{k})x)$ at large distances in real spaces. Thus, only the QNMs with an exponentially damped field along the trajectory in the complex plane are transformed into square integrable wavefunctions. This happens only if $\exp(i\tilde{k}\tilde{X}) \to 0$ when $\tilde{X} \to \infty + i\infty$. For a vacuum, $\tilde{k} = \widetilde{\omega}/c$, the necessary condition is met only for QNMs with a quality factor $Q = -\frac{1}{2}\text{Re}(\widetilde{\omega})/\text{Im}(\widetilde{\omega})$ that verifies

$$\tan(\theta) > 1/(2Q). \qquad (12)$$

Therefore, in the mapped space of regularized QNMs, only a subset of the initial complex QNMs emerges [Via14]. Figure 3(a) which is directly inspired from Fig. 2 in [Via14] illustrates our purpose. Only the QNMs with large quality factors emerge in the new complex frequency plane of the mapped space. They have complex frequencies that are identical to the initial ones obtained with an unscaled calculation, see Fig. 1. In the next Section 3.3, we will clarify what is happening with the veiled QNMs in the hatched half plane.

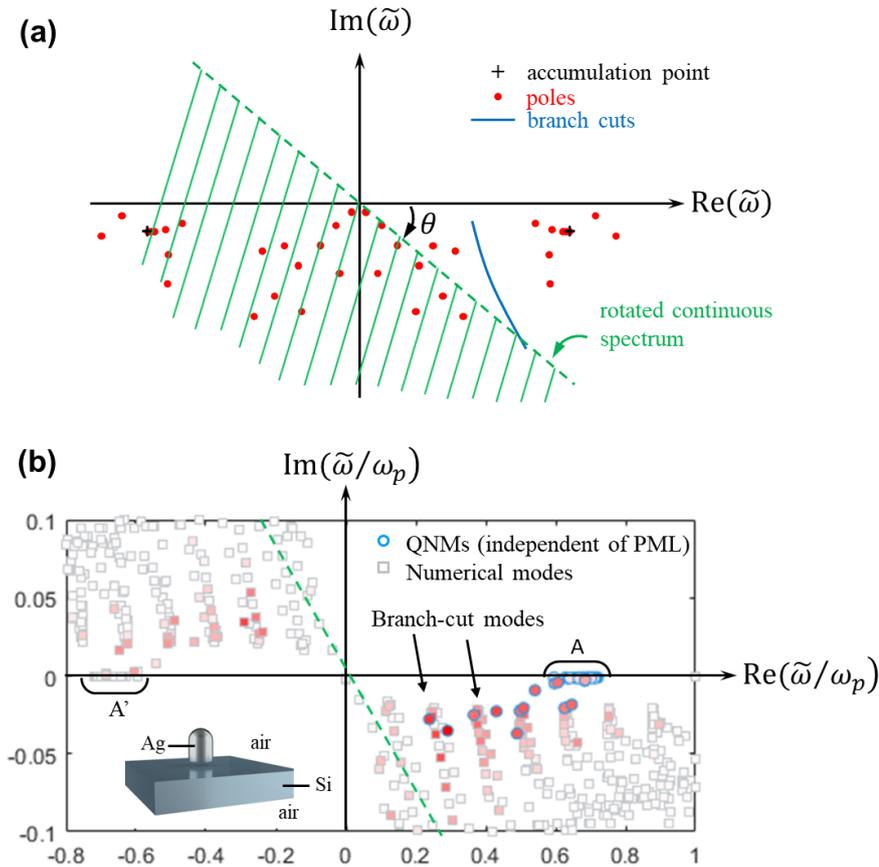

Fig. 3. **Spectrum of regularized spaces.** (a) Complex frequency plane after regularization by a non-dispersive PML with an infinite thickness. Owing to the rotation in the complex plane, the continuous spectrum composed of propagative or evanescent radiation modes (the real axis in the initial unmapped space) is rotated by an angle $\theta$. Its rotated version is shown with the dashed green line. The latter divides the plane in two half planes. Above the line, the initial QNMs are unveiled by the regularization and the spectrum is unchanged. Below the line in the shaded half plane, the initial QNMs are veiled and the spectrum is largely impacted. Inspired by Fig. 2 in [Via14]. (b) Complex-frequency plane of a discretized operator for a quite complicated geometry, a Drude-silver nanorod (plasma frequency $\lambda_p \equiv 2\pi c/\omega_p$ =138 nm, 60-nm diameter, 90-nm height) lying on a 138-nm-thick Si slab. Several branch cuts are discretized; two of them are marked with arrows. Modes corresponding to QNMs are shown with a circle. They are independent of the PML. Numerical modes are shown with square marks. The red color indicates the excitation rate of the modes (here upon illumination by a plane wave), dark and light red corresponding to highly and weakly excitations, respectively. The spectrum was computed with the QNMEig solver using a non-dispersive PML. After Fig. SI.7(a) in [Yan18].

*Importantly,* note that the regularized QNM fields $\widetilde{\Psi} = [\widetilde{\mathbf{E}}; \widetilde{\mathbf{H}}]$ are different from the original diverging fields that are solutions of Eq. (1). They only coincide in the unscaled interior segment $|x| < x_0$. Another *important* remark concerns the fact that the field of the QNM in the real (physical) space $x$ for $|x| > x_0$ cannot be recovered from the QNM field calculated in the PML for $|x'| > x_0$. That would be possible only if a one-to-one mapping would exist between $x$ and $x'$. And this does not happen since a PML is a branch cut in the complex plane which helps satisfying the outgoing wave condition. De facto, the one-to-one mapping only exists between the complex trajectory $\tilde{X} = F(x')$ and $x'$. It is thus important to realize that the PML regularization does not implement a similarity transform between the initial divergent QNMs and the regularized ones (the operator **U** of Eq. (2) is meaningless in the present context). The PML machinery that generally applies to a wide range of geometries and not just to analytical functions is a much more abstract concept than the simplified similarity transform approach introduced see the Section 3.1.

We know from Section 2 that the entire set of QNMs provides completeness only on rare occasions. Now we see that only an infinite subset of QNMs are unveiled by the regularization. Thus, the question of the completeness of regularized QNMs arises. That is where the numerical modes come in, as we discuss see the Sections 3.4.3 and 3.4.4. Let us first illustrate what are these numerical modes.

### 3.3 Practical PML-regularization with finite and discretized spaces

Except for some analytical cases, such as 1D Fabry-Perot resonators or spherical Mie potentials, the complex mapping with an infinite spatial extend is truly inconvenient. In practice, QNMs are computed with numerical methods [Lal19], e.g., differential methods, integral methods, modal methods. The numerical space is always finite and owing to the 'discretization', the operators become matrices with a finite size. Numerical PMLs with finite thicknesses are similar to the PMLs introduced in the previous section. They additionally feature perfectly conducting magnetic or electric walls on the outer boundaries of the PML space (Neumann or Dirichlet boundary condition for a finite value $x' > x_0$), so that the mapped space becomes bounded. To account for this change, we symbolically rewrite the regularized spectral problem of Eq. (11) as

$$\nabla \times \widetilde{\mathbf{E}}(\mathbf{r}^B) = i\widetilde{\omega}\boldsymbol{\mu}'(\mathbf{r}^B, \widetilde{\omega})\widetilde{\mathbf{H}}(\mathbf{r}^B),$$
$$\nabla \times \widetilde{\mathbf{H}}(\mathbf{r}^B) = -i\widetilde{\omega}\boldsymbol{\varepsilon}'(\mathbf{r}^B, \widetilde{\omega})\widetilde{\mathbf{E}}(\mathbf{r}^B), \quad (13)$$
+ hard wall conditions at the outer PML boundary.

In Eq. (13), the superscript '$B$' holds for 'bounded'. It will be omitted hereafter for simplicity reasons.

Indeed, since the regularized QNMs are exponentially decaying in the PML region, most of them are weakly impacted by truncating the mapped space by hard wall: only their tiny exponential tails "see" the wall and reflect off it, before being attenuated again *on the way back* towards the region of interest $x' < x_0$. However, some QNMs have small quality factors and are diverging extremely fast. Their regularized field are not damped enough before reaching the hard wall and are quite impacted by the hard-wall boundaries. Thus, the spectrum of the new bounded operator with hard walls may be quite different from the one with infinite PMLs in Eq. (11). However, importantly, the actual QNMs that are relevant to understand the physics of the resonators in some frequency range of interest are hopefully preserved by the regularization and the space truncation.

Figure 3 illustrates the significant impact on the spectrum of the passage from a continuous operator in an unbounded space to a discretized one in a bounded space. The spectrum corresponds to a silver nanorod on a thin Si layer. The spectrum was computed with the solver QNMEig operating under COMSOL environment with non-dispersive PMLs, see details in [Yan18].

The spectrum encompasses physical resonances, i.e., the QNMs shown with blue circles, and numerical modes shown with squares. For this complex geometry, the numerical modes and the QNMs set close to one another in the complex-frequency plane. To discriminate them, the authors in [Yan18] have exploited the fact that the QNMs are insensitive to PML-parameter variations (as long as they are revealed by the complex coordinate transform, see the Section 3.4), whereas numerical modes are sensitive [note7]. They have thus performed two computations for two different PML parameters and have identified the eigenvalues that are independent of the PML. Further details are provided in the Supplemental Section 3.4 in [Yan18].

The second salient feature is that the spectrum includes several nearly vertical aligned rows of numerical modes which are regularly spaced along the real axis. The PML implements the outgoing wave conditions for guided modes of the slab, and just like for free-space plane waves in uniform media, differentiating ingoing and outgoing guided modes requires a cut in the complex plane. This cut is automatically implemented by the PML (at least for some specific geometries where the phase velocity is opposite to the group velocity [note5]) and the aligned rows of numerical modes are thus interpreted as *discretized* branch cuts operated by the different waveguide modes [Mar91]. Note that similar (discretized) branch patterns are also observed for gratings due to the passing-off of diffraction orders at cutoff frequencies determined by the grating equation, see Fig. 11 in [Via14] and Fig. 1 in [Gra19].

Some numerical modes, especially those with a negative frequency ($\text{Re}(\widetilde{\omega}) < 0$), have positive imaginary parts ($\text{Im}(\widetilde{\omega}) > 0$), implying that they exponentially diverge as $t \to \infty$. The unphysical modes (we are not considering amplifying media) are due to the fact that the PML used for the computation are non-dispersive and therefore does not satisfy the Hermitian symmetry [note8]. Note that they do not prevent reaching a high accuracy when reconstructing the field scattered in the temporal domain by a driving pulse. Indeed as $t \to \infty$, the reconstruction becomes false, but energy has leaked away from the resonator well before divergence effects become observable, see Fig. 2(b) in [Yan18] and the related discussion.

In Fig. 3, all the modes are colored and sorted by increasing order of their excitation coefficient $\alpha_m$ in Eq. (8). Dark red coloring corresponds to a strong impact. QNMs are mostly impacting, but we also note some numerical modes arising from the branch cuts which significantly contribute to the reconstruction. This evidences that the branch cuts cannot be neglected in general, and that leakage into membrane mode is a dominant channel in this particular example.

Further note that some numerical modes shown with a square, like those labelled by the letter $A'$, are nearly independent of the PML but are not considered as QNMS since the imaginary part of their frequency is positive owing to the use of non-dispersive PMLs. They are disembodiment of the Hermitian-symmetric QNMs labelled by the letter $A$.

### 3.4 Fundamental properties of regularized QNMs

*3.4.1 Norm of regularized QNMs*

The norm of regularized QNMs relies on a volume integral of the form $\int \left[ \widetilde{\mathbf{E}} \cdot \frac{\partial \omega \boldsymbol{\varepsilon}}{\partial \omega} \widetilde{\mathbf{E}} - \widetilde{\mathbf{H}} \cdot \frac{\partial \omega \boldsymbol{\mu}}{\partial \omega} \widetilde{\mathbf{H}} \right] d^3 \mathbf{r}$, see the Section 4 and [Sau13]. The integral is undefined for the unscaled real space owing to the QNM divergence, but it is defined for the mapped space for the regularized QNMs that are square integrable since they are exponentially damped in the mapped space for large $\mathbf{r}$. The question arises as to whether the integral over the infinite regularized space depends on the complex mapping (or PML) used?

The key point is that the integral, and therefore the norm, is independent of the trajectory used, as long as the inequality in Eq. (12) is satisfied, indeed. Since the integral over the complex trajectory is equal to the integral over the mapped space $x$' (the demonstration that relies on the equivalence between material parameters and spatial coordinate transforms [Nic94] is not repeated here but can be found in the Supplementary Section 2 in [Sau13]), the PML norm computed over the entire space is independent of the PML considered. Appendix 2 provides another demonstration of this key property. The demonstration is very simple and can be followed by all the readership.

The norm is thus a unique quantity, which is intrinsic to the QNM and leads to an effective and quite general method to compute normalized QNMs in practice [Sau13,Yan18,Lal19], see the Section 3 for a comparison of the PML-norm with other rigorous norms.

*3.4.2 Biorthogonality*

The concept of orthogonality is essential, and its demonstration is quite trivial for regularized QNMs. We will consider the QNMs of nanoresonators built with non-dispersive materials for the sake of simplicity. For dispersive materials, the permittivity depends on the frequency and Eq. (11) corresponds to a nonlinear eigenvalue problem. To derive the orthogonality relation, one needs to linearize the operator [Lal18]. This can be performed with auxiliary fields [Ram10,Yan18]. In general, the auxiliary fields are physical quantities that represent the material polarisations **P** and currents **J** and the biorthogonality relation holds between QNMs enlarged with the auxiliary fields, $\widetilde{\boldsymbol{\Psi}} = [\widetilde{\mathbf{E}}; \widetilde{\mathbf{H}}; \widetilde{\mathbf{P}}; \widetilde{\mathbf{J}}]$, which combine geometrical and material resonances [note9].

The following derivation holds for non-dispersive materials [Sau13]. It is repeated here to evidence that regularized spaces and their regularized eigenvectors can be treated in much the same way that the normal modes of Hermitian system are treated. For that, please compare the following derivation with that of Section entitled "General Properties of the Harmonic Modes" in [Joa08].

Let us consider two regularized QNMs $\widetilde{\boldsymbol{\Psi}}_n = [\widetilde{\mathbf{E}}_n; \widetilde{\mathbf{H}}_n]$ and $\widetilde{\boldsymbol{\Psi}}_m = [\widetilde{\mathbf{E}}_m; \widetilde{\mathbf{H}}_m]$, therein satisfying the same source-free Maxwellian operator, Eq. (11). We simply apply the Lorentz reciprocity theorem, taking the QNM $\widetilde{\boldsymbol{\Psi}}_n$ with a complex-valued frequency $\widetilde{\omega}_n$ for solution 1 and the QNM $\widetilde{\boldsymbol{\Psi}}_m$ with a complex-valued frequency $\widetilde{\omega}_m$ for solution 2. Since there is no source, we obtain

$$\int_\Sigma \{\widetilde{\mathbf{E}}_m \times \widetilde{\mathbf{H}}_n - \widetilde{\mathbf{E}}_n \times \widetilde{\mathbf{H}}_m\} \cdot \mathbf{n} dS = i(\widetilde{\omega}_n - \widetilde{\omega}_m) \int_\Omega \{\widetilde{\mathbf{E}}_n \cdot \boldsymbol{\varepsilon} \widetilde{\mathbf{E}}_m - \widetilde{\mathbf{H}}_n \cdot \boldsymbol{\mu} \widetilde{\mathbf{H}}_m\} dV, \qquad (14)$$

an equation that is valid for any surface $\Sigma$ delimitating the compact subspace $\Omega$. The orthogonality relation is obtained by considering the entire space. Since the regularized QNMs asymptotically vanish for $r \to \infty$, the surface integral tends to zero and we have $(\widetilde{\omega}_n - \widetilde{\omega}_m) \int_{\Omega \cup \Omega_{\text{PML}}} \{\widetilde{\mathbf{E}}_n \cdot \boldsymbol{\varepsilon} \widetilde{\mathbf{E}}_m - \widetilde{\mathbf{H}}_n \cdot \boldsymbol{\mu} \widetilde{\mathbf{H}}_m\} dV = 0$. For $n \neq m$ and non-degenerate modes, the volume integral is equal to zero, and if normalized, the regularized QNMs of non-dispersive systems are biorthogonal [note10]

$$\int_{\Omega \cup \Omega_{\text{PML}}} \{\widetilde{\mathbf{E}}_n \cdot \boldsymbol{\varepsilon} \widetilde{\mathbf{E}}_m - \widetilde{\mathbf{H}}_n \cdot \boldsymbol{\mu} \widetilde{\mathbf{H}}_m\} dV = \delta_{nm}. \qquad (15)$$

If two QNMs have equal frequencies $\widetilde{\omega}_n = \widetilde{\omega}_m$, then they are degenerate and are not necessarily orthogonal. For two modes to be degenerate requires that two different field patterns happen to have precisely the same frequency. Usually, there is a symmetry that is responsible for the astonishing coincidence. For example, if the permittivity and permeability distributions are invariant under a 90° rotation, QNMs that differ only by a 90° rotation are expected to have the same frequency. Such QNMs are degenerate and are not necessarily orthogonal. However, since the Maxwellian operator is linear, any linear combination of these degenerate QNMs is itself a mode with that same frequency. We can always choose to work with linear combinations that are orthogonal [Gra20].

*3.4.3 Completeness of expansions including numerical modes*

So far, we have examined the completeness of expansions composed of 'true' QNMs and argued that completeness is achieved for very limited geometries only. In this Section, we examine the critical issue of the completeness of expansions involving regularized QNMs and numerical modes in 'PMLized' spaces.

The theoretical case of an unbounded space with infinitely thick PML has been scarcely considered in the literature to our knowledge. Little studies have been carried out to understand what happens to the veiled QNMs below the rotated continuous spectrum, whether they are replaced by new poles, and what is the nature of these poles that are not 'true' QNMs. It would be interesting to determine the entire spectrum of mapped operators in infinite spaces and possibly their branch cuts [Via14,Oly04].

In fact, most recent studies have focused on practical cases, i.e. on mapped spaces with finite-thickness PMLs bounded by perfectly conducting electric or magnetic walls and discretized operators (i.e. non-Hermitian matrices). The theoretical background, essentially linear algebra, is much simpler than for continuous operators defined on open spaces. In contrast to a Hermitian matrix, a non-Hermitian matrix does not have an orthogonal set of eigenvectors; in other words, a non-Hermitian matrix $A$ can in general not be transformed by an orthogonal matrix $Q$ to diagonal form $D = Q^*AQ$, where the star is the transpose conjugate. However, most non-Hermitian matrices can be transformed by a nonorthogonal matrix $X$ to diagonal form $D = X^{-1}AX$ through their right and left eigenvectors [note11]. There exist rare exceptions, for which $X$ converges to a singular operator. This case corresponds to defective matrices is often referred to as non-Hermitian degeneracy or self-orthogonality, see for instance the Chapter 9 in [Moi11] and [El18].

Except for these exceptional cases for which two or more complex eigenvalues cross and the associated eigenvectors coalesce, the literature generally assumes that the whole set of QNMs and numerical modes (also termed as PML modes or Bérenger modes) form a complete set. This has been demonstrated for problems for which the Green's tensor is analytically known [Oly04]. However, although no rigorous mathematical proof formally exists, the conjecture is supported by many numerical examples.

Though they play a key role in the analysis of non-Hermitian matrices [Arf05], left eigenvectors (or left QNMs) have not been considered so far in this review. For instance, only right QNMs appear in the biorthogonality product of Eq. (15). The reason is that we have considered so far only reciprocal materials ($\boldsymbol{\varepsilon} = \boldsymbol{\varepsilon}^T$ and $\boldsymbol{\mu} = \boldsymbol{\mu}^T$ where the superscript 'T' stands for transpose). In this case of broad practical interest, the right and left QNMs (or eigenvectors) are equal and it is therefore not appropriate to distinguish them. For further reading, please refer to [Zol18,Zha20] and Appendix C, where the normalization of QNMs of resonators with non-reciprocal materials is considered.

Concerning purely numerical issues, the approach adopted in [Sau13] is similar (the Lorentz reciprocity normalization and the right-left eigenvector normalizations are equivalent [Gra20]). For dispersive media, the eigenproblem is no longer linear and the right-left eigenvector orthonormalization becomes impossible. In [Sau13], it is shown with the Lorentz reciprocity theorem that normalization is still possible but that the lack of orthogonality leads to an absence of analyticity for the modal coefficients of the QNM expansion, a fact numerically evidenced in Fig. 8(a) in [Lal18].

Convincingly, accurate reconstructions of the scattered field with QNM expansions augmented by numerical modes were first provided for a 2D dielectric triangular rod, for which convergence of the absorption cross-section spectrum was reported by increasing the total number of QNMs and numerical modes retained in the expansion, see Fig. 7 in [Via14].

For resonators made of dispersive materials, the eigenvalue problem defined in Eq. (13) is no longer linear, since the operator itself depends on the frequency via $\boldsymbol{\varepsilon}(\widetilde{\omega})$ or $\boldsymbol{\mu}(\widetilde{\omega})$. A possible approach to recover linearity is to incorporate auxiliary-fields, for instance, the material polarisations $\mathbf{P}$ and currents $\mathbf{J}$. It is then possible to recover a biorthogonal relation (see Eq. (4) in [Yan18]) for both QNMs and numerical modes. The biorthogonality can then be used for projection on the basis elements andan expression for the excitation coefficients, the $\alpha$'s, can be further derived. Note that, like for any Hermitian matrix, once the source term is defined, there is a unique closed-form expression for the $\alpha$'s per biorthogonal relation [Yan18]. However, as shown in [Gra20], new formulas can be found by choosing different auxiliary fields for linearization or different methods to spitting the source term [note12].

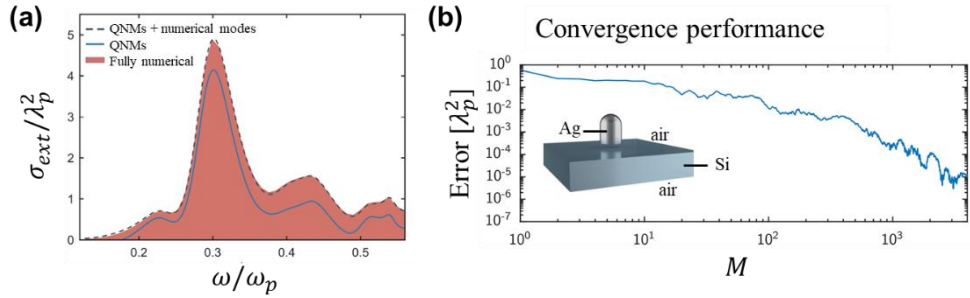

**Fig. 4 Numerical evidence of the completeness of expansions with QNMs and PML modes**. (a) Scattering cross-section spectrum of the Ag nanobullet of Fig. 3(a). The solid-blue and dashed-black curves are computed by retaining 20 QNMs only and 20 QNMs plus 200 numerical modes, respectively, and the shadowed pink curves are the data obtained with the frequency-domain solver of COMSOL Multiphysics. After [Yan18]. (b) Convergence rate as the total number $M$ of modes retained in the double-sum expansion in Eq. (8) is increased. The vertical axis represents the spectrally-averaged $(0.1\omega_p < \omega < 0.6\omega_p)$ error between the predictions of the extinction cross section obtained with the truncated modal expansion of Eq. (8) and fully-vectorial data computed with COMSOL Multiphysics. The results are obtained with discretized spaces regularized with finite PML thicknesses and the modes are sorted by increasing order of their impact on the reconstruction. Adapted from Fig. SI.7 in [Yan18]. Other relevant examples illustrating completeness can be found in [Via14,Zol18,Gra19,Gra20].

Figure 4 shows the convergence of the extinction cross-section spectra of the Ag nanorod for a broad range of frequency of six octaves. Despite the series of branch cuts and accumulation points (at the pole and zero of the Drude-silver permittivity and at the surface plasmon frequency), convergence is evidenced as the total number $M$ of numerical modes and QNMs retained in the expansion of Eq. (8) is increased. The reference [Yan18] contains other numerical evidence of the convergence for complicated geometries, such as the accurate reconstruction of light scattered by bowtie antennas in the temporal domain with short pulses or quenched emission at frequency close to the surface plasmon frequency and its accumulation point.

Other convincing reconstructions have also been obtained for other complicated geometries, e.g metallic gratings that also have accumulations points and multiple branch cuts at grating anomalies [Gra19]. In [Zol18], the convergence of the expansion is tested for a 2D resonator with a prismatic shape and a Lorentz permittivity for an even broader spectral interval, though a small inaccuracy is reported close to the pole of the permittivity with $M = 200$ modes. In [Gra20], the authors have considered simpler geometries, e.g. 2D Drude cylinder in free space. Using dispersive PMLs, they have been able to compute all the QNMs and have numerically evidenced that the double-sum expansion is complete, inside or outside the cylinder, i.e. everywhere in the physical domain that corresponds to the unmapped subspace [note13]. The authors have also checked that expansions based on QNMs is incomplete also inside the resonator, owing to branch-cut of the 2D Green function. See Fig. 2 in [Gra20] for more details.

The second line in Table 1 summarizes the discussion.

### 3.5 Multiphysics

The regularization approach with mapped spaces can be applied to problems combining electromagnetism and other areas of physics. First, it is worth noting that, for dispersive materials, multiphysics is already enrolled in the regularization approach with auxiliary fields. This is because the dispersions of $\varepsilon$ and $\mu$ are determined by the microscopic interaction between electromagnetic fields and other elementary excitations, such as phonons, polaritons, magnons, plasmons [Riv20].

For dispersive materials, the permittivity or permeability depend on the frequency and Eq. (1) corresponds to a non-linear eigenvalue problem. One then encounters a difficulty in deriving an orthogonality relation, which prevents us to obtain a closed-form expression for the modal excitation coefficient of the scattered fields ($\mathbf{E}_s$, $\mathbf{H}_s$), as the excitation coefficient of one QNM depends on the other QNMs [Sau13]. This issue can be elegantly solved by introducing auxiliary fields which allow casting the non-linearly eigenvalue into a linear one [Yan18]. For example, considering a medium with a dispersive permittivity described by a single-pole Lorentz model, $\varepsilon(\omega) = \varepsilon_\infty - \varepsilon_\infty \omega_p^2 (\omega^2 - \omega_0^2 + i\omega\gamma)^{-1}$, two auxiliary fields, the polarization $\mathbf{P} = -\varepsilon_\infty \omega_p^2 (\omega^2 - \omega_0^2 + i\omega\gamma)^{-1} \mathbf{E}$ and the current density $\mathbf{J} = -i\omega \mathbf{P}$, can be introduced to reformulate Eq. (1) into a higher dimensional *linear* eigenvalue problem. An orthogonality relation, $\iiint_{\Omega \cup \Omega_{PML}} \varepsilon_\infty \tilde{\mathbf{E}}_n \cdot \tilde{\mathbf{E}}_m - \mu_0 \tilde{\mathbf{H}}_n \cdot \tilde{\mathbf{H}}_m + \omega_0^2/(\varepsilon_\infty \omega_p^2) \tilde{\mathbf{P}}_n \cdot \tilde{\mathbf{P}}_m - 1/(\varepsilon_\infty \omega_p^2) \tilde{\mathbf{J}}_n \cdot \tilde{\mathbf{J}}_m d^3\mathbf{r} = 0$ for $m \neq n$, can then be obtained by incorporating the auxiliary fields $\mathbf{P}$ and $\mathbf{J}$ into the Lorentz reciprocity theorem. With this orthogonality relationship, it is possible to project the scattered fields ($\mathbf{E}_s, \mathbf{H}_s, \mathbf{P}_s, \mathbf{J}_s$) into the enlarged QNM vector ($\tilde{\mathbf{E}}_m, \tilde{\mathbf{H}}_m, \tilde{\mathbf{P}}_m, \tilde{\mathbf{J}}_m$), which encompasses electromagnetic fields and material properties. This projection leads to the derivation of a closed-form expression for the QNM excitation coefficient [Yan18].

For more complicated cases involving cross-action of Maxwell's resonant field with other equations of physics, e.g., optomechanical cooling [Kip08], lasing [Tur06], photonic switching [Sha19], the physics of the system can no longer be fully taken into account with the medium response functions $\boldsymbol{\varepsilon}$ and $\boldsymbol{\mu}$. The norms and orthogonality relations need to be rederived. In general, the derivation can be performed by strictly following the method in [Sau13,Yan18]. The most important thing one needs to care about is to fully account for the contributions of the extra current distributions, modified polarization fields, or boundary conditions, when using the Lorentz reciprocity theorem.

An illustrative example is found with QNMs incorporating non-classical response functions to accurately model electronic spill-out, nonlocality, or surface-enabled Landau damping for deep subwavelength confinements. This can be achieved thanks to an effective nonclassical surface polarization, $\mathbf{P}_{\text{surf}} = \boldsymbol{\pi} + i\omega^{-1}\mathbf{K}$ with $\boldsymbol{\pi}$ being an out-of-plane dipole moment and $\mathbf{K}$ an in-plane surface current density, treated with the Feibelman $d$-parameters formalism [Fei82]. Interestingly, it has been found that, in addition to a volume integral term, the normalization formulas acquire a surface contribution arising from the nonclassical surface polarization [Zho21]. Another important thing worth mentioning is that, for dispersive $d$ parameters, the self-consistent Maxwell's equation defines a non-linear eigenvalue problem, even when $\boldsymbol{\varepsilon}$ and $\boldsymbol{\mu}$ are frequency-independent. In analogy to the case of dispersive permittivities mentioned above, the Maxwell's equations can be again mapped into a higher dimensional linear eigenvalue problem by introducing a surface auxiliary field, and thus an orthogonality relationship can be readily obtained [Zho21].

## 4. Normalization of QNMs

It is interesting to go through the history of QNM normalization with the theoretical concepts presented see the Section 3 in mind. Two distinct periods can be conveniently considered. In the 1990's, P. T. Leung, K. Young, and coworkers at the Chinese University of Hong Kong played an important role in the development of QNM theory. They studied scalar fields in 1D cavities as well as vector fields in 3D systems with a spherical symmetry. Except for a few works that extended these results to 1D photonic bandgap structures [Set03,Set09], the domain remained dormant for 10 years. Then, the 2010's witnessed a second active period where different groups addressed the issue of QNM normalization for arbitrary 3D geometries for which no analytical solution of Maxwell's equations are available.

Figure 5 summarizes the main achievements of these two periods and places them along a timeline. Four different methods have been elaborated in the 2010's to normalize QNMs of 3D optical resonators made of dispersive materials. The methods have already been compared [Ge14,Kri15,Sau15,Mul16a,Mul17,Kri17] and a few numerical tests have also been performed

[Kri15,Sau15,Mul16a]. However, no clear consensus has been reached. Some points have even been vividly debated in a series of articles, comment and reply [Kri15,Mul16a,Mul17,Kri17]. Some works consider that three integral expressions of the norm nearly provide the same result [Kri15,Kri20] and that a difference is noticeable only on rare occasions. On the contrary, some other works have shown that significant differences exist [Sau15,Mul16a,Mul17] and have argued that the normalization inherited from the work by the Hong-Kong group (referred to as LK normalization hereafter) is incorrect [Mul16a,Mul17]. In addition, some arguments raised in [Ge14,Kri15] concerning the PML normalization are questionable [Lal20].

For all these reasons, we believe that it is important to compare in detail the different norms, so that future works can be built on firm ground. Hereafter, we start by a historical perspective presenting the evolution of ideas, then we review previous arguments and results exposed in [Ge14,Kri15,Mul16a,Mul17,Kri17], to further provide new mathematical demonstrations and numerical evidences. We hope that the whole clarifies the issue.

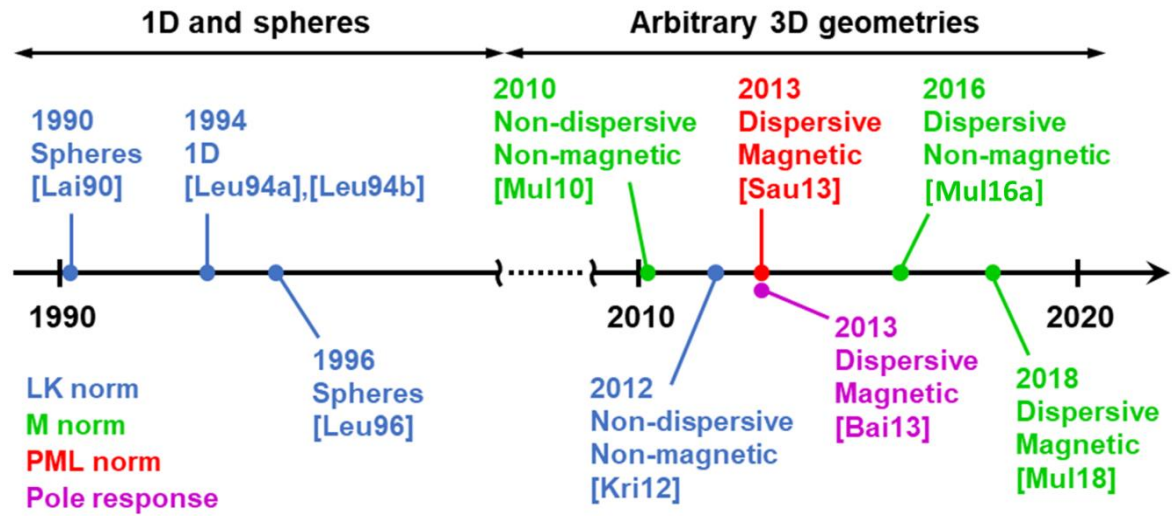

Fig. 5. Milestones in the evolution of ideas on QNM normalization in electromagnetism. Two different periods can be identified. Normalization of scalar fields in 1D cavities and vector fields in 3D systems with a spherical symmetry have mainly been studied during the 1990's. The 2010's witnessed a second active period on QNM normalization for arbitrary 3D geometries. The emphasis was also placed during this period on the numerical implementation. The four normalization methods presented see the Section 4 are distinguished by different colors: blue (LK norm), green (M norm), red (PML norm), and magenta (pole response). The adjectives "non-dispersive", "dispersive", "non-magnetic", and "magnetic" refer to the material properties considered in the corresponding works.

### 4.1 Historical perspective on QNM normalization
*4.1.1 The 1990's: 1D and spherical cavities*
The 1990's witnessed important development in QNM theory, essentially with the works from the Chinese University of Hong Kong [Lai90,Leu94a,Leu94b,Leu96,Lee99]. The Hong Kong group derived important theorems on the orthogonality and the completeness of QNM expansions for scalar fields in 1D cavities. In [Leu94a], QNMs are regularized by considering an analytic continuation for complex coordinates, exactly as discussed see the Section 3. The 1D regularization allowed them to define the QNM norm and the inner product from which results the QNM orthogonality. In addition, they showed that QNMs form a complete set inside the cavity, essentially with the arguments developed see the Section 2.3. However, the PML machinery was not yet invented at that time and the QNM regularization approach has not been further developed for other geometry than the 1D Fabry-Perot cavity.

During the course of their studies, the Hong Kong researchers have also developped another approach to normalize the QNMs of dielectric microspheres [Lai90]. They presented their approach as an extension of the Zel'dovich 1D regularization formalism [Zel61] to vector fields in 3D. However, no regularization is implemented and the norm depends on a volume integral and a surface integral. We refer to this norm as the Leung-Kristensen (LK) norm, using the same terminology as in [Mul16a,Mul17]. For dispersive but non-magnetic ($\mu = \mu_0$) media, it can be written as

$$\langle \widetilde{\boldsymbol{\Phi}}_m | \widetilde{\boldsymbol{\Phi}}_m \rangle_{\text{LK}} = \lim_{\Omega \to \infty} \iiint_\Omega \widetilde{\mathbf{E}}_m \cdot \left( \varepsilon + \frac{\partial \omega \varepsilon}{\partial \omega} \right) \widetilde{\mathbf{E}}_m d^3 \mathbf{r} + I_{\text{surf}}^{\text{LK}}, \tag{16}$$

where the derivative with respect to the frequency is taken at $\omega = \widetilde{\omega}_m$ and $I_{\text{surf}}^{\text{LK}}$ is a surface integral over the outer border Σ of the compact domain Ω,

$$I_{\text{surf}}^{\text{LK}} = \frac{i\varepsilon_0 nc}{\widetilde{\omega}_m} \oiint_\Sigma \widetilde{\mathbf{E}}_m^2 dS, \tag{17}$$

with $\varepsilon_0$ the vacuum permittivity, $n$ the refractive index of the uniform medium surrounding the resonator and $c$ the speed of light. Note that the factor that takes into account the permittivity dispersion can be written slightly differently due to the equality $\varepsilon + \partial(\omega\varepsilon)/\partial\omega = 2\partial(\omega^2\varepsilon)/\partial(\omega^2)$.

Let us emphasize that the limit $\Omega \to \infty$ was not present in the initial proposal [Lai90] and that the authors were aware of the lack of rigour of their norm; they recognized in Appendix A that the norm is accurate only for QNMs with a large quality factor and for a given range of sizes of the domain Ω surrounding the resonator.

The limit $\Omega \to \infty$ has been introduced later on in [Leu96], and was popularized in a series of works in the 2010's [Kri12,Kri14b,Ge14,Kri15] that studied cavities with non-spherical shapes, e.g., photonic-crystal cavities. Taking the limit suggests that the sum of the volume and surface integrals, which depend on the size of Ω, becomes constant when the size of the integration domain Ω is asymptotically large. In agreement with [Mul16a,Mul17], we show see the Sections 4.2.2 and 4.2.3 that this suggestion is fully incorrect.

To understand the 2010's developments and debates, it is worth coming back on the two proposals of the Hong Kong group: the norm with the complex-coordinate regularization (integration along a path in the complex plane) and the LK norm with a surface integral term. For 1D resonators, the scalar field outside the cavity is simply given by an outgoing plane wave with a complex propagation constant. The integral in the complex plane starting at $x = R$ can thus be calculated analytically and it is simply proportional to $E_m^2(R)$, the QNM field squared at $x = R$, see the Section IV.A in [Lee09]. Both norms are thus strictly equal for 1D cavities. *However, in 2D and 3D, the field outside the cavity is no longer a plane wave and the equivalence between the two norms is broken.* Therefore, the surface term used in [Lai90,Leu96,Lee99] is incorrect for 3D systems (even spheres). This point was already mentioned as a tricky issue in Appendix A of [Lai90] but, unfortunately, it has been forgotten in subsequent works.

*4.1.2 The 2010's: towards arbitrary 3D geometries*

The turn of the twenty-first century witnessed major progress of nanotechnologies and it became possible to fabricate photonic structures with an increasingly complex geometry. Consistently, high demand prompted the development of QNM theories for 3D arbitrary geometries for which no analytical solution of Maxwell's equations is known. The 2010's have been particularly active and fruitful with the contributions of different groups.

**LK norm.** Let us start by the work of P. T. Kristensen and colleagues who used the normalization proposed in the 1990's, initially for computing the mode volume of high-Q photonic-crystal cavities [Kri12] and then plasmonic antennas [Ge14,Kri14b], for which the validity of the LK norm has rapidly raised questions and mistrust.

**M norm.** A different norm was available since the beginning of the decennia. In 2010, E. A. Muljarov, W. Langbein, and R. Zimmermann were developing a Brillouin-Wigner perturbation theory for open electromagnetic systems, which rapidly became known as the resonant-state expansion (RSE) [Mul10]. They derived a QNM norm in the form of a sum of a volume integral and a surface integral, initially for non-dispersive and non-magnetic materials [Mul10], then for dispersive materials [Mul16a], and finally for magnetic (even bi-anisotropic) materials [Mul18].

In the following, we refer to this normalization method as the M norm. In its most general form, it can be written as [Mul18]

$$\langle \widetilde{\boldsymbol{\phi}}_m | \widetilde{\boldsymbol{\phi}}_m \rangle_\text{M} = \iiint_\Omega \left( \widetilde{\mathbf{E}}_m \cdot \frac{\partial \omega \boldsymbol{\varepsilon}}{\partial \omega} \widetilde{\mathbf{E}}_m - \widetilde{\mathbf{H}}_m \cdot \frac{\partial \omega \boldsymbol{\mu}}{\partial \omega} \widetilde{\mathbf{H}}_m \right) d^3\mathbf{r} + I_\text{surf}^\text{M}, \qquad (18)$$

where the spectral derivatives are taken at $\omega = \widetilde{\omega}_m$ and $I_\text{surf}^\text{M}$ is a surface integral over the outer border Σ of the domain Ω,

$$I_\text{surf}^\text{M} = \frac{i}{\widetilde{\omega}_m} \oiint_\Sigma \left[ \widetilde{\mathbf{E}}_m \times (\mathbf{r} \cdot \nabla) \widetilde{\mathbf{H}}_m - (\mathbf{r} \cdot \nabla) \widetilde{\mathbf{E}}_m \times \widetilde{\mathbf{H}}_m \right] \cdot d\mathbf{S}. \qquad (19)$$

The derivation of Eqs. (18)-(19) relies on the divergence theorem, an analytic continuation of the QNM field to a function of complex frequencies, and a Taylor expansion close to the QNM eigenfrequency [Doo14,Mul16a,Mul18]. First-order spectral derivatives are then replaced by spatial derivatives of the QNM field under the assumption that the resonator is embedded in a uniform background. Note that slightly different expressions of the M norm have been proposed over the years. The initial expression in [Mul10] contains first and second-order spatial derivatives of the electric field. The M norm can also be formulated with only first-order spatial derivatives of the electric field, as shown in Appendix B of [Mul16a]. Finally, the most general expression of the M norm given by Eqs. (18)-(19) contains first-order spatial derivatives, only, of both the electric and magnetic fields. Albeit technical, these last remarks are important to further understand the domain of validity, the ease of numerical implementation (see Appendix D), and the accuracy of the computed norms (see the Section 4.2.4).

At first sight, the M norm, Eqs. (18)-(19), and the LK norm, Eqs. (16)-(17), appear similar: they both encompass a surface integral to compensate the divergence of the volume integral as the size of the integration domain Ω is enlarged. However, the surface integrals are different. It has recently been shown that the surface term $I_\text{surf}^\text{LK}$ is incorrect, whereas the surface term $I_\text{surf}^\text{M}$ provides correct normalization for resonators embedded in a homogeneous background (no substrate) [Mul16a,Mul17]. These key points are confirmed by simulations results and mathematical evidence see the Section 4.2.

**PML norm.** In 2013, C. Sauvan and his colleagues proposed to work with PML-regularized QNMs [Sau13]. As discussed see the Section 3, the PML normalization is an incarnation of the generalized version (for vector fields in 3D) of the complex coordinate transform introduced in quantum mechanics for simple systems and scalar fields. We refer to this normalization method as the PML norm. The latter relies on the sum of two volume integrals [Sau13]

$$\langle \widetilde{\boldsymbol{\phi}}_m | \widetilde{\boldsymbol{\phi}}_m \rangle_\text{PML} = \iiint_\Omega \left( \widetilde{\mathbf{E}}_m \cdot \frac{\partial \omega \boldsymbol{\varepsilon}}{\partial \omega} \widetilde{\mathbf{E}}_m - \widetilde{\mathbf{H}}_m \cdot \frac{\partial \omega \boldsymbol{\mu}}{\partial \omega} \widetilde{\mathbf{H}}_m \right) d^3\mathbf{r} + I_\text{PML}, \qquad (20)$$

where the spectral derivatives are taken at $\omega = \widetilde{\omega}_m$ and $I_\text{PML}$ is a volume integral with the same integrand over the PML domain $\Omega_\text{PML}$ that surrounds the unmapped domain Ω,

$$I_\text{PML} = \iiint_{\Omega_\text{PML}} \left( \widetilde{\mathbf{E}}_m \cdot \frac{\partial \omega \boldsymbol{\varepsilon}}{\partial \omega} \widetilde{\mathbf{E}}_m - \widetilde{\mathbf{H}}_m \cdot \frac{\partial \omega \boldsymbol{\mu}}{\partial \omega} \widetilde{\mathbf{H}}_m \right) d^3\mathbf{r}. \qquad (21)$$

The permittivity and permeability in Eq. (21) are those inside the PML area, see the Section 3.

It is interesting to note that the first term of the PML norm, Eq. (20), is identical to the first term of the M norm, Eq. (18). Therefore, the PML norm and the M norm are identical if, and only if, the surface integral $I_\text{surf}^\text{M}$ is equal to the volume integral $I_\text{PML}$ in the PMLs, $I_\text{PML} = I_\text{surf}^\text{M}$. See the Section 4.2.4, we prove this equality for resonators embedded in a uniform background (no substrate).

PML-regularized QNMs have been studied in further works. As we have seen see the Section 3.3, PML-regularization is also well suited to the computer world of linear algebra problems with huge matrices of *finite* dimensions, after discretization of the continuous operator. Normalization is then ensured by biorthogonal products and the use of right and left eigenvectors, see Chapter 6 in [Moi11]. When the right and left eigenvectors are computed with a numerical technique that uses PMLs, the PML normalization is then naturally implemented [Röm08,Via14,Gra20]. Note that for reciprocal materials, the left and right eigenvectors are identical and numerical normalization due to biorthogonal products of matrices corresponds to a discretized version of the PML norm in finite space, as evidenced in [Gra20]. Further note that the biorthogonality relation obtained for discretized operators and finite spaces also corresponds to the orthogonality relation derived for regularized QNM in [Sau13] for reciprocal materials.

**Pole-response norm.** Finally, let us introduce a fourth normalization method that was proposed in [Bai13]. It is quite different from the three others since it does not rely on the calculation of integrals of the QNM field. In contrast, it relies on the calculation of the electromagnetic field excited by a monochromatic point source located inside or in the vicinity of the resonator. The normalization relies on the fact that, for *complex* frequencies $\omega$ sufficiently close to the complex eigenfrequency $\widetilde{\omega}_m$ of the QNM of interest, the radiated field is quasi identical to the QNM field. It is then possible to derive an analytical expression for the response of the system driven at a complex frequency close to the eigenfrequency, see [Bai13] and Section 4.2 in [Lal18]. We refer to this method as the pole-response normalization.

The pole-response normalization is a very general method. It led to the freeware QNMPole [QNMPole] launched in 2013. It can be used with any frequency-domain electromagnetic solver (using PMLs or not), with any dispersive materials (as long as analytical expressions of the frequency-dependent permittivity and permeability are known), and for "any" geometries including microcavities that leak into semi-infinite periodic waveguides. The latter situation, which can be often encountered in photonic-crystal integrated optics, was successfully tested in [Fag17].

## 4.2 Exactness and domain of validity of the different norms

We test the four different norms for various geometries, highlight their domain of validity and mathematically demonstrate equivalence between the M norm and the PML norm for the specific case of uniform backgrounds. The Section combines results available from the literature and new results, to fully clarify the normalization issue.

### *4.2.1 Two methods available for almost any geometries*

Before testing all the normalization methods, let us remind that the LK and M norms can only be applied to resonators embedded in a uniform background. They are not valid for geometries such as resonators deposited on a substrate (e.g., nanoparticles on a mirror [Aks14,San16], photonic crystal cavities [Lal08]), embedded in a thin film stack [Fah09], or coupled to a waveguide [Tan07,Fag17]. In contrast, the PML normalization and the pole-response normalization remain valid for these geometries. Note, however, that the PML norm cannot be used for cavities coupled to a periodic waveguide, because of the lack of PMLs that fulfill outgoing wave boundary conditions in a periodic waveguide that supports Bloch modes instead of a usual waveguide that supports translation-invariant guided modes. The pole-response normalization remains valid in that case [Fag17].

Figure 6 compares the PML normalization and the pole-response normalization for a resonator deposited on a metal substrate . The 2D geometry is depicted in Fig. 6(a). It consists of a thin dielectric patch (refractive index 1.5, thickness 30 nm and length 100 nm) sandwiched between a gold substrate and a gold patch (thickness 30 nm and length 100 nm). We have calculated the lowest-order magnetic dipolar mode for transverse magnetic (TM) polarization (magnetic field polarized along the $z$-direction). Its electric-field modulus is displayed in Fig. 6(b).

Maxwell's equations are solved with the aperiodic Fourier Modal Method (a-FMM) [Sil01], by retaining 2001 Fourier harmonics in the Fourier expansion of the electromagnetic fields along the $x$- direction. For the evaluation of the PML norm, the PMLs have been implemented with uniform layers of anisotropic and absorbing materials. The corresponding permittivity and permeability tensors ($\mu_z, \varepsilon_x, \varepsilon_z$ in TM polarization) are equal to the permeability and permittivity of the medium in which the PML is added (air or metallic substrate, see Fig. 6) multiplied by the complex factor $f_{\mathrm{PML}}$ or $1/f_{\mathrm{PML}}$ depending on the tensor component [Hug05]. We have chosen the thickness and the complex factor of the PML ($h_{\mathrm{PML}} = 2\mathrm{Re}(\tilde{\lambda}_m)/3$ and $f_{\mathrm{PML}} = 15 + 15i$) so that the ratio between the modal field at the outer PML boundaries and the maximum field is smaller than $10^{-5}$. The same PML parameters have been used for all values of $R$. We have checked that different parameters provide the same result (within the numerical accuracy) provided that the field is very weak at the outer PML boundaries.

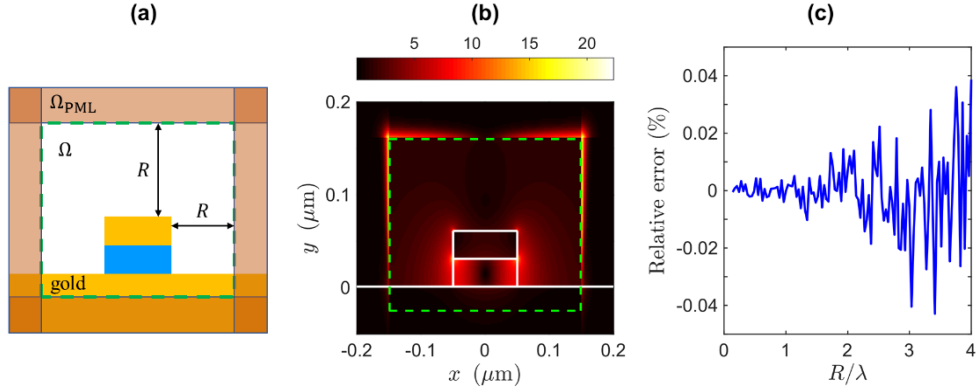

**Fig. 6.** Comparison between the PML normalization and the pole-response normalization for a 2D resonator on a metal substrate. (a) Sketch of the nanopatch resonator together with a definition of the integration domains $\Omega$ and $\Omega_{\mathrm{PML}}$ used to normalize the mode with Eq. (20). (b) Distribution of the modulus of the normalized electric field of the lowest-order magnetic dipolar mode ($\tilde{\lambda}_m = 725 + i54$ nm). The dashed lines represent the PML boundaries. Note the damping of the field inside the PMLs. (c) Relative error on the real part between the PML norm and the pole-response norm, $\mathrm{Re}\left(\langle\tilde{\boldsymbol{\Phi}}_m|\tilde{\boldsymbol{\Phi}}_m\rangle_{\mathrm{PML}} - \langle\tilde{\boldsymbol{\Phi}}_m|\tilde{\boldsymbol{\Phi}}_m\rangle_{\mathrm{pole}}\right)/\mathrm{Re}\left(\langle\tilde{\boldsymbol{\Phi}}_m|\tilde{\boldsymbol{\Phi}}_m\rangle_{\mathrm{pole}}\right)$. Adapted from [Sau15].

Figure 6(b) shows the relative error between the PML norm and the pole-response norm as the size $R$ of the computational domain $\Omega$ is increased. Since the pole-response norm does not depend on $R$, we have chosen to calculate it for the smallest computational domain for the sake of numerical precision. The PML norm is also independent of discretization, theoretically. However, as $R$ is increased, the discretization grain decreases since the number of Fourier harmonics is constant, and numerical errors increase. An irrelevant tiny variation ($< 0.04\%$) of the norm, due to numerical inaccuracies, is observed as $R$ is increased by a factor 30, from $R = 0.13\lambda$ to $4\lambda$. To our knowledge, this calculation is the only comparison between different normalization methods for the important practical case of a resonator that is not embedded in a homogeneous medium.

*4.2.2 Stringent difference between the M and LK norms*
In this Section, we discuss a key claim about the LK, M and PML norms (see the abstract in [Kri15]): "*We discuss three formally different formulas for normalization of quasinormal modes currently in use for modeling optical cavities and plasmonic resonators and show that they are complementary and provide the same result*". The statement is simple and comforting, but it is unfortunately wrong. As demonstrated with an analytical treatment for dielectric spheres in air in [Mul16a,Mul17], the LK and M norms provide different results. In Fig. 7, we further support this point for a metallic (dispersive) sphere. In addition, we compare the LK and M norms with the PML norm.

We consider the electric dipolar QNM of lowest order supported by a silver sphere (radius 40 nm) in air. We compute the PML, LK and M norms with known analytical expressions of the QNM field for spheres [Doo14]. The volume and surface integrals are calculated numerically with the Gaussian quadrature rule. The integration domain $\Omega$ is a sphere of radius $R$ and the calculation is repeated for increasing values of $R$. Since the QNM field takes a simple analytical expression in spherical coordinates, the PML is implemented as a radial complex mapping inside a 4-μm-thick shell surrounding the unmapped domain Ω defined by $r < R$, $r \to R + (1 + i/2)(r - R)$ for $R < r < R + 4$ μm.

As $R$ is increased from 60 nm to 4 μm, the PML norm remains constant with an excellent accuracy (variations smaller than $5 \times 10^{-13}$). This value is further used as a reference to test the M and LK norms. Figure 7 displays the relative errors (in %) on the real parts of the LK norm, $\text{Re}\left(\langle \tilde{\boldsymbol{\phi}}_m | \tilde{\boldsymbol{\phi}}_m \rangle_{\text{LK}} - \langle \tilde{\boldsymbol{\phi}}_m | \tilde{\boldsymbol{\phi}}_m \rangle_{\text{PML}}\right) / \text{Re}\left(\langle \tilde{\boldsymbol{\phi}}_m | \tilde{\boldsymbol{\phi}}_m \rangle_{\text{PML}}\right)$ (solid blue curve), and M norm, $\text{Re}\left(\langle \tilde{\boldsymbol{\phi}}_m | \tilde{\boldsymbol{\phi}}_m \rangle_{\text{M}} - \langle \tilde{\boldsymbol{\phi}}_m | \tilde{\boldsymbol{\phi}}_m \rangle_{\text{PML}}\right) / \text{Re}\left(\langle \tilde{\boldsymbol{\phi}}_m | \tilde{\boldsymbol{\phi}}_m \rangle_{\text{PML}}\right)$ (dashed red curve).

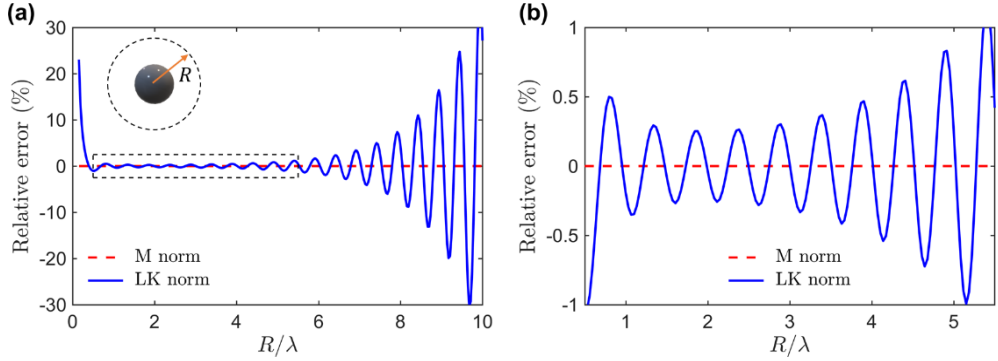

**Fig. 7.** Comparison between the PML, LK and M norms for lowest-order electric dipolar mode of a silver sphere of radius 40 nm ($\tilde{\lambda}_m = 390 + i32$ nm). (a) The PML norm is used as a reference value to benchmark the relative error on the real parts of the L norm, $\text{Re}\left(\langle \tilde{\boldsymbol{\phi}}_m | \tilde{\boldsymbol{\phi}}_m \rangle_{\text{LK}} - \langle \tilde{\boldsymbol{\phi}}_m | \tilde{\boldsymbol{\phi}}_m \rangle_{\text{PML}}\right) / \text{Re}\left(\langle \tilde{\boldsymbol{\phi}}_m | \tilde{\boldsymbol{\phi}}_m \rangle_{\text{PML}}\right)$ (solid blue curve), and the M norm, $\text{Re}\left(\langle \tilde{\boldsymbol{\phi}}_m | \tilde{\boldsymbol{\phi}}_m \rangle_{\text{M}} - \langle \tilde{\boldsymbol{\phi}}_m | \tilde{\boldsymbol{\phi}}_m \rangle_{\text{PML}}\right) / \text{Re}\left(\langle \tilde{\boldsymbol{\phi}}_m | \tilde{\boldsymbol{\phi}}_m \rangle_{\text{PML}}\right)$ (dashed red curve). (b) provides an enlarged view of the dashed rectangle in (a). The sphere (radius 40 nm) is embedded in air. The permittivity of silver is given by a Drude model taken from [Arc10]. The PML in spherical coordinates is implemented as a finite-size complex coordinate transform between $R$ and $R + 4$ μm, $r \to R + (1 + i/2)(r - R)$. For the M norm, the first-order derivatives in $I_{\text{surf}}^{\text{M}}$ are calculated with a centered finite-difference scheme.

Three conclusions concerning the LK normalization are beyond question: (1) the LK norm is different from the PML and M norms, (2) the LK norm is not independent on the size $R$ of the calculation domain, and (3) the LK norm diverges as $R$ increases. The relative error is quite large even for small domains ($> 1\%$ for $R < \lambda/2$ and even $> 10\%$ for $R < \lambda/5$). It oscillates around zero for $R > \lambda/2$. The amplitude of the oscillation first decreases before exhibiting an exponential increase. The smallest error (0.3%) is obtained for $R \approx 2\lambda$. The reason for the divergence of the LK norm is analyzed see the Section 4.2.3 and in [Mul16a,Mul17].

See the Section 4.2.4, we mathematically prove that, for resonators embedded in a uniform medium, the M and PML norms are equivalent. Thus, we are not surprised that the M and PML norms provide identical results in Fig. 7. A closer inspection to the data reveals a relative deviation between the two norms that is smaller than $10^{-4}\%$. The deviation is due to difficulties encountered for the numerical implementation of the M norm, as discussed in Appendix D.

*4.2.3 Divergence of the LK norm*
The fact that the LK norm fails to solve the issue of the divergence of the QNM field has already been documented in [Kri15,Mul16a,Mul17]. For instance, Fig. 3 in [Kri15] shows the same result as Fig. 7, i.e., an oscillation with an increasing amplitude as the size of the integration domain is increased. For the simple case of a sphere, the LK norm can be derived analytically. Equations (24)-(26) in [Kri15] and

Eq. (C15) in [Mul16a] unambiguously evidence that the LK norm diverges as $\exp(Re(\widetilde{\omega}_m)R/Qc)$, with $Q$ the quality factor of the mode.

We now show that the LK norm is, in fact, an approximation of the first term of the PML norm given by Eq. (20). It means that the LK norm completely misses the regularization contribution due to the volume integral in the PML. The surface term $I_{\text{surf}}^{\text{LK}}$ in Eq. (16) is actually a decoy: it does not compensate the divergence of the volume integral over the domain $\Omega$, but simply partly accounts for the volume integral of the magnetic field.

By using the divergence theorem and radiation conditions on the border $\Sigma$ of a large spherical domain $\Omega$, we show in Appendix E that

$$I_{\text{surf}}^{\text{LK}} \approx - \iiint_\Omega \left(\widetilde{\mathbf{E}}_m \cdot \boldsymbol{\varepsilon} \widetilde{\mathbf{E}}_m + \widetilde{\mathbf{H}}_m \cdot \boldsymbol{\mu} \widetilde{\mathbf{H}}_m\right) d^3\mathbf{r} \tag{22}$$

The $\approx$ sign simply comes from the use of radiation conditions, which are not rigorous for a finite-size domain $\Omega$. As a result, we find that a good approximation for the LK norm is

$$\langle \widetilde{\boldsymbol{\phi}}_m | \widetilde{\boldsymbol{\phi}}_m \rangle_{\text{LK}} \approx \iiint_\Omega \left(\widetilde{\mathbf{E}}_m \cdot \frac{\partial \omega \boldsymbol{\varepsilon}}{\partial \omega} \widetilde{\mathbf{E}}_m - \widetilde{\mathbf{H}}_m \cdot \boldsymbol{\mu} \widetilde{\mathbf{H}}_m\right) d^3\mathbf{r}. \tag{23}$$

This is exactly the expression of the first term of the PML norm in Eq. (20) for non-dispersive magnetic materials. Equation (23) clearly shows that $\langle \widetilde{\boldsymbol{\phi}}_m | \widetilde{\boldsymbol{\phi}}_m \rangle_{\text{LK}}$ is actually an integral of the QNM field over the domain $\Omega$ only. The absence of any regularization simply explains why the LK norm exponentially grows as $R \to \infty$.

Let us conclude by commenting the arguments developed in [Kri15] in favor of considering the LK norm as valid and useful, despite the inconsistency of its divergence. See the conclusion in [Kri15]: "*The surface term in the normalization by Lai et al. represents a simple choice of regularization which in principle is insufficient to properly regularize the integral when the size of the calculation domain is varied along the real axis. We have discussed this issue in detail, and we have shown how one can, in principle, always regularize the integral by a complex coordinate transform. In practice, we find that there is rarely any need for additional regularizations beyond the relatively simple formula of Lai et al.*" Arguably, the divergence is not a real problem since the LK norm can be regularized to avoid the divergence. We agree: the LK norm can be regularized, but this is precisely what is doing the PML norm or the M norm for resonators in free space (see next Section). It is important to be aware that the LK norm by itself, as proposed in [Lai90,Leu96,Kri12,Kri14b,Ge14], does not contain this regularization and is thus incorrect. Directly adopting a mathematically correct approach, instead of hoping numerical errors to alleviate fundamental issues, avoids misleading discussions and flaws inevitably echoed by others colleagues.

*4.2.4 Equivalence between the PML and M norms for resonators in free space*
The first term of the PML norm, see Eq. (20), is identical to the first term of the M norm, see Eq. (18). Therefore, the PML norm and the M norm provide the same result if, and only if, $I_{\text{PML}}$ in Eq. (21) and $I_{\text{surf}}^{\text{M}}$ in Eq. (19) are also identical. In the following, we show that $I_{\text{PML}} = I_{\text{surf}}^{\text{M}}$ for resonators surrounded by an uniform medium (no substrate).

For the demonstration, we follow the same approach as in [Mul16a,Mul18]. We consider the field radiated by a source emitting at a complex frequency $\omega$ that tends towards the QNM frequency $\widetilde{\omega}_m$ and further use the Lorentz reciprocity theorem to transform the volume integral in $I_{\text{PML}}$ into the surface integral in $I_{\text{surf}}^{\text{M}}$.

Let us consider the electromagnetic field $\boldsymbol{\phi} = (\mathbf{E}, \mathbf{H})$ solution of Maxwell's equations at the frequency $\omega$ for an electric current source $\mathbf{j}(\mathbf{r}, \omega) = (\omega - \widetilde{\omega}_m)\mathbf{S}(\mathbf{r})$ located in the domain $\Omega$. Because of the frequency dependence of the current, $(\mathbf{E}, \mathbf{H}) \to (\widetilde{\mathbf{E}}_m, \widetilde{\mathbf{H}}_m)$ as $\omega \to \widetilde{\omega}_m$. Note that the mode $\widetilde{\boldsymbol{\phi}}_m$ is not necessarily normalized.

We apply the Lorentz reciprocity theorem (see Appendix A) in the domain $\Omega_{\text{PML}}$ to the field $\boldsymbol{\phi}$ at the frequency $\omega$ and to the mode $\widetilde{\boldsymbol{\phi}}_m$ at $\widetilde{\omega}_m$. Since it is located inside $\Omega$, the source $\mathbf{j}(\mathbf{r}, \omega)$ is located outside $\Omega_{\text{PML}}$, the source term in Eq. (A2) cancels out. Additionally, since $\widetilde{\boldsymbol{\phi}}_m$ and $\boldsymbol{\phi}$ vanish at the outer border of $\Omega_{\text{PML}}$, the surface term reduces to an integral over the closed surface $\Sigma$. After dividing each term by $(\omega - \widetilde{\omega}_m)$, we obtain

$$i \iiint_{\Omega_{\text{PML}}} \mathbf{E} \cdot \frac{\omega \boldsymbol{\varepsilon}(\omega) - \widetilde{\omega}_m \boldsymbol{\varepsilon}(\widetilde{\omega}_m)}{\omega - \widetilde{\omega}_m} \widetilde{\mathbf{E}}_m - \mathbf{H} \cdot \frac{\omega \boldsymbol{\mu}(\omega) - \widetilde{\omega}_m \boldsymbol{\mu}(\widetilde{\omega}_m)}{\omega - \widetilde{\omega}_m} \widetilde{\mathbf{H}}_m d^3 \mathbf{r}$$
$$= - \oiint_{\Sigma} \frac{1}{\omega - \widetilde{\omega}_m} \left[ \widetilde{\mathbf{E}}_m \times \mathbf{H} - \mathbf{E} \times \widetilde{\mathbf{H}}_m \right] \cdot d\mathbf{S}, \quad (24)$$

where the vector $d\mathbf{S}$ is the outward pointing unit vector at each point of the surface $\Sigma$. We now take the limit $\omega \to \widetilde{\omega}_m$. Since $(\mathbf{E}, \mathbf{H}) \to (\widetilde{\mathbf{E}}_m, \widetilde{\mathbf{H}}_m)$, we easily see that the left-hand term tends towards the volume integral $I_{\text{PML}}$. For the limit of the right-hand term, the indeterminate form can be evaluated by considering a Taylor expansion of the field $\boldsymbol{\phi} = (\mathbf{E}, \mathbf{H})$

$$\boldsymbol{\phi} = \widetilde{\boldsymbol{\phi}}_m + (\omega - \widetilde{\omega}_m) \frac{\partial \boldsymbol{\phi}}{\partial \omega} + o(\omega - \widetilde{\omega}_m). \quad (25)$$

Finally, taking the limit $\omega \to \widetilde{\omega}_m$ in Eq. (24) provides

$$\iiint_{\Omega_{\text{PML}}} \left( \widetilde{\mathbf{E}}_m \cdot \frac{\partial \omega \boldsymbol{\varepsilon}}{\partial \omega} \widetilde{\mathbf{E}}_m - \widetilde{\mathbf{H}}_m \cdot \frac{\partial \omega \boldsymbol{\mu}}{\partial \omega} \widetilde{\mathbf{H}}_m \right) d^3 \mathbf{r} = i \oiint_{\Sigma} \left[ \widetilde{\mathbf{E}}_m \times \frac{\partial \mathbf{H}}{\partial \omega} - \frac{\partial \mathbf{E}}{\partial \omega} \times \widetilde{\mathbf{H}}_m \right] \cdot d\mathbf{S}, \quad (26)$$

where all derivatives with respect to the frequency are taken at $\omega = \widetilde{\omega}_m$.

So far, we have not made any assumption on the medium surrounding the resonator and Eq. (26) shows that it is possible to replace the volume integral over the PML domain $\Omega_{\text{PML}}$ in Eq. (20) by a surface integral over its inner boundary $\Sigma$. However, without any further assumption that we detail hereafter, the left-hand term, and thus the M norm, does not only rely on the QNM itself but additionally encompasses the derivative with respect to the frequency, $\frac{\partial \mathbf{E}}{\partial \omega}$ and $\frac{\partial \mathbf{H}}{\partial \omega}$, of the analytic continuation of the QNM.

This issue can be solved if we assume that the resonator is surrounded by a homogeneous medium. In that case, the fields outside the resonator consist of only outgoing waves; they are functions of the variable $\omega \mathbf{r}$ and the spectral derivatives can be rewritten as [Mul16a]

$$\left. \frac{\partial \mathbf{E}}{\partial \omega} \right|_{\omega = \widetilde{\omega}_m} = \frac{1}{\widetilde{\omega}_m} (\mathbf{r} \cdot \nabla) \widetilde{\mathbf{E}}_m, \quad (27)$$

which holds for a field $\widetilde{\mathbf{E}}_m$ consisting of only outgoing or only incoming waves propagating in a homogeneous medium surrounding the system. A similar relation holds for the magnetic field. Equation (26) finally becomes

$$I_{\text{PML}} = \frac{i}{\widetilde{\omega}_m} \oiint_{\Sigma} \left[ \widetilde{\mathbf{E}}_m \times (\mathbf{r} \cdot \nabla) \widetilde{\mathbf{H}}_m - (\mathbf{r} \cdot \nabla) \widetilde{\mathbf{E}}_m \times \widetilde{\mathbf{H}}_m \right] \cdot d\mathbf{S} = I_{\text{surf}}^{\text{M}}. \quad (28)$$

Conclusively, for resonators embedded in a homogeneous medium, $I_{\text{PML}} = I_{\text{surf}}^{\text{M}}$ and the PML and M normalizations, respectively in [Sau13] and in [Mul18a] are strictly equivalent.

**Table 1**. Summary of normalization methods used in the 2010's for reciprocal materials[a].

|  |  | **PML norm** [Sau13] Eqs. (20)-(21) | **M norm** [Mul18][b] Eqs. (18)-(19) | **Pole-response norm** [Bai13] |
|---|---|---|---|---|
| **Principle** |  | Integral over the entire regularized space, $\Omega \cup \Omega_{\text{PML}}$ | Integral over a finite compact volume $\Omega$ | Response to an excitation at a complex frequency $\widetilde{\omega} \approx \widetilde{\omega}_m$ |
| **Regularization** |  | Volume integral inside the PMLs | Surface integral over the boundary of $\Omega$ | Not necessary |
| **Materials** | **Dispersive** | ✓ | ✓ | ✓ |
|  | **Magnetic** | ✓ | ✓ | ✓ |
|  | **Bianisotropic** | ✗[c] | ✓ | ✓ |

| Background | Homogeneous medium | ✓ | ✓ | ✓ |
|---|---|---|---|---|
| | Substrate or multilayer stack | ✓ | X | ✓ |
| | Waveguide | ✓ | X | ✓ |
| | Periodic waveguide | X | X | ✓ |

[a]The derivation of the QNM norm for non-reciprocal materials is not an issue (see Appendix D). We have chosen to not to present in the Table since the norms have not been compared. Note that the LK norm is not documented in the Table because it is inexact even for simple geometries, e.g. dielectric spheres in air.
[b]Initially proposed in [Mul10] for non-dispersive and non-magnetic materials.
[c]The PML norm can be straightforwardly generalized to bianisotropic media. However, the generalization is unpublished.

## 5. Conclusion

In this didactic article, we have reviewed key concepts attached to the quasinormal modes of open electromagnetic systems: the normalization, orthogonality, and completeness. Normalization allows us to directly quantify the strength of the interaction of light with a resonance. Orthogonality enables the projection of any field on the QNM basis. Completeness guaranties the convergence and exactness of QNM expansions of the scattered field.

The normalization of QNMs in electromagnetism has not been achieved correctly and an error or approximation introduced in the 90's has been unfortunately repeated and amplified in the 10's, leading to some confusion in the literature. The difficulty is due to the divergence of the QNM field in the open space. We would again emphasize that the divergence is not unphysical. The reverse is actually true. We have clarified the confusion, putting forth the errors made and quantifying the domain of validity of the correct methods.

Nowadays the normalization is no longer a theoretical issue. Methods and even free softwave are available since almost ten years to normalize the QNMs of (virtually) any system. Perhaps, a relevant question is to derive numerical methods that compute normalized QNMs in a very effective and accurate way. And there is room for imagination. In the same spirit of the pole-response normalization with a source, we may envision illuminating the resonator with a plane wave at a complex frequency close to the eigenfrequency of the QNM of interest, and the scattered field will be dominated by a single QNM. The normalization factor can be readily deduced from the scattered field and closed-form expressions of the excitation coefficient $\alpha_m$ [note14]. Other options may rely on the use of analytical expression available on the first-order QNM perturbation theory.

We have recalled a worthwhile result, quite well known for a long time [Bau71,Bau76, Mor71,Leu94a], on the completeness of expansions based on true QNMs. Completeness is restrictively warranted inside the resonators only and additionally for resonators surrounded by free space (no substrate). Hopefully, *approximate* or truncated expansions, relying on a few dominant resonant modes only, may already be very useful for experiment interpretations or initial designs before optimization takes place. However, completeness is an enjoyable and reassuring mathematical property to guaranty convergence and exactness of various formalism, e.g. high-order QNM perturbation theory or QNM coupled mode theory.

We have highlighted a natural and quite general way to complement the incomplete QNM to achieve completeness: QNM regularization. The latter has been widely considered in quantum mechanical studies with quite different approaches [Moi11]. We have described recent results in electromagnetism, aiming at setting up effective formalisms and numerical tools based on complex coordinate mapping, viewed as PMLs. We have presented data to show that an extended basis composed of regularized QNMs and additional numerical modes forms a complete basis of the regularized Hilbert spaces. Interestingly, we may use orthonormality relations that are quite similar to those used in "Hermitian electrodynamics" [Har61] and many geometries of current interest in plasmonics and nanophotonics can be faithfully analyzed.

After 30 years or more, research on QNMs in electromagnetism continues unabated. This reflects the underlying importance of resonances in the contemporary physics and applications. It seems very likely that this situation will continue. We expect that the review contributes to contextualize and clarify recent achievements obtained the last decennia.

## 6. Acknowledgements

PL would like to thank Wei Yan for his multiple contributions to the present work and analysis.

## Appendix A. Unconjugated form of Lorentz reciprocity theorem for absorbing and dispersive media

Consider a system characterized by the position- and frequency-dependent permittivity and permeability tensors $\boldsymbol{\varepsilon}(\mathbf{r}, \omega)$ and $\boldsymbol{\mu}(\mathbf{r}, \omega)$. We assume reciprocal materials, $\boldsymbol{\mu} = \boldsymbol{\mu}^T$ and $\boldsymbol{\varepsilon} = \boldsymbol{\varepsilon}^T$, where the superscript $T$ denotes matrix transposition. The electromagnetic field $(\mathbf{E}, \mathbf{H})$ radiated by a current source density $\mathbf{j}$ at the frequency $\omega$ is given by the two curl Maxwell's equations,

$$\nabla \times \mathbf{E} = i\omega\boldsymbol{\mu}(\boldsymbol{r}, \omega)\mathbf{H} \quad \text{and} \quad \nabla \times \mathbf{H} = -i\omega\boldsymbol{\varepsilon}(\mathbf{r}, \omega)\mathbf{E} + \mathbf{j}. \tag{A1}$$

Lorentz reciprocity theorem relates two different solutions of Maxwell's equations, $(\mathbf{E}_1, \mathbf{H}_1, \omega_1, \mathbf{j}_1)$ and $(\mathbf{E}_2, \mathbf{H}_2, \omega_2, \mathbf{j}_2)$ labelled by the index 1 and 2. It is derived by applying the divergence theorem to the vector $\mathbf{E}_2 \times \mathbf{H}_1 - \mathbf{E}_1 \times \mathbf{H}_2$ and by using Eq. (A1),

$$\iint_\Sigma (\mathbf{E}_2 \times \mathbf{H}_1 - \mathbf{E}_1 \times \mathbf{H}_2) \cdot \mathbf{n} dS$$
$$= i \iiint_\Omega \{\mathbf{E}_1[\omega_1\boldsymbol{\varepsilon}(\omega_1) - \omega_2\boldsymbol{\varepsilon}(\omega_2)] \cdot \mathbf{E}_2 - \mathbf{H}_1[\omega_1\boldsymbol{\mu}(\omega_1) - \omega_2\boldsymbol{\mu}(\omega_2)] \cdot \mathbf{H}_2\} dV$$
$$- \iiint_\Omega (\mathbf{j}_1 \cdot \mathbf{E}_2 - \mathbf{j}_2 \cdot \mathbf{E}_1) dV, \tag{A2}$$

where $\Sigma$ is an arbitrary closed surface defining a volume $\Omega$ and $\mathbf{n}$ is the unit vector normal to the surface $\Sigma$ (oriented outward).

## Appendix B. Direct proof of the invariance of the PML-norm

The following demonstration is very general and quite straightforward. Let us consider a resonator composed of non-magnetic materials and illuminated by a source at frequency $\omega$. In the scattered field formulation, the scattered field $[\mathbf{E}_s, \mathbf{H}_s]$ satisfies the Maxwell's equations with a current distribution inside the resonator $\mathbf{J}(\omega) = -i\omega\Delta\boldsymbol{\varepsilon}(\mathbf{r}, \omega)\mathbf{E}_b(\omega, \boldsymbol{r})$, where $\Delta\boldsymbol{\varepsilon}(\mathbf{r}, \omega)$ is null everywhere except in the resonator and $\mathbf{E}_b(\omega, \boldsymbol{r})$ is the driving field. All details with the same notations are found in Annex 2 in [Lal18].

We consider a first regularized version of the scattering problem, in which a PML is introduced outside the resonator that necessarily has a finite spatial extend ($\Delta\boldsymbol{\varepsilon}$ is null outside a compact support $\Omega_{res}$). Referring to Fig. 2, this defines an unmapped compact domain bounded by the inner boundary of the PML domain, a rotational angle $\theta$ that satisfies Eq. (12) and a stretching coefficient if any.

We assume that the $m^{th}$ QNM is revealed by the PML and denote its regularized field by $[\widetilde{\mathbf{E}}_m, \widetilde{\mathbf{H}}_m]$. By applying the Lorentz reciprocity theorem, see Appendix A, we find

$$\int_{\Omega \cup \Omega_{\text{PML}}} [\mathbf{E}_s \cdot (\omega \boldsymbol{\varepsilon}(\omega) - \widetilde{\omega}_m \boldsymbol{\varepsilon}(\widetilde{\omega}_m))\tilde{\mathbf{E}}_m - (\omega \boldsymbol{\mu}(\omega) - \widetilde{\omega}_m \boldsymbol{\mu}(\widetilde{\omega}_m))\mathbf{H}_s \cdot \tilde{\mathbf{H}}_m] dV = -i \int_\Omega \mathbf{J} \cdot \tilde{\mathbf{E}}_m dV, \quad (B1)$$

where the integral is performed over the entire space, including the unmapped compact domain $\Omega \supseteq \Omega_{res}$ and the PML domain $\Omega_{\text{PML}}$. Note that the surface-integral term vanishes and does not show up in Eq. (B1) because all fields, $\mathbf{E}_s, \tilde{\mathbf{E}}_m,\ldots$ exponentially vanish for $|\mathbf{r}| \to \infty$. QNMs are poles and thus, the scattered field diverges for any $\mathbf{r}$ as the source frequency $\omega \to \widetilde{\omega}_m$. For a pole of order 1, the residue can be written as

$$\lim_{\omega \to \widetilde{\omega}_m} (\omega - \widetilde{\omega}_m)[\mathbf{E}_s(\mathbf{r}, \omega), \mathbf{H}_s(\mathbf{r}, \omega)] = \gamma_m [\tilde{\mathbf{E}}_m(\mathbf{r}), \tilde{\mathbf{H}}_m(\mathbf{r})]. \quad (B2)$$

Similar considerations apply to poles of higher orders. $\gamma_m$ is a complex coefficient that may be simply determined by evaluating the limit of Eq. (B1) as $\omega \to \widetilde{\omega}_m$. By using Eq. (B2) and the definition of the derivative at $\omega = \widetilde{\omega}_m$, $\lim_{\omega \to \widetilde{\omega}_m} [\omega \boldsymbol{\varepsilon}(\omega) - \widetilde{\omega}_m \boldsymbol{\varepsilon}(\widetilde{\omega}_m)]/(\omega - \widetilde{\omega}_m) \equiv \frac{\partial \omega \boldsymbol{\varepsilon}}{\partial \omega}$, we obtain [Sau13]

$$\gamma_m = \frac{-i \int_{\Omega_{res}} \mathbf{J}(\widetilde{\omega}_m) \cdot \tilde{\mathbf{E}}_m dV}{\int_{\Omega \cup \Omega_{\text{PML}}} \left[ \tilde{\mathbf{E}}_m \cdot \frac{\partial \omega \boldsymbol{\varepsilon}}{\partial \omega} \tilde{\mathbf{E}}_m - \mu_0 \tilde{\mathbf{H}}_m \cdot \tilde{\mathbf{H}}_m \right] dV}. \quad (B3)$$

Note that the source term is given by $\mathbf{J}(\widetilde{\omega}_m) = -i\widetilde{\omega}_m \Delta \boldsymbol{\varepsilon}(\mathbf{r}, \widetilde{\omega}_m)\mathbf{E}_b(\widetilde{\omega}_m, \mathbf{r})$.

Now let us consider another regularization performed with a new PML that is different from the previous one, possibly because the new unmapped compact domain $\Omega'$ or the new rotational angle are different. To demonstrate the invariance of the PML norm, we assume that the QNM $[\tilde{\mathbf{E}}_m, \tilde{\mathbf{H}}_m]$ is again revealed by the new PML, i.e., the new rotational angle $\theta'$ satisfies Eq. (12). The new QNM, denoted $[\tilde{\mathbf{E}}'_m, \tilde{\mathbf{H}}'_m]$, is identical to the previous one in the unmapped space: $[\tilde{\mathbf{E}}'_m, \tilde{\mathbf{H}}'_m] = [\tilde{\mathbf{E}}_m, \tilde{\mathbf{H}}_m]$ for $\mathbf{r} \in \Omega \cap \Omega'$. It differs from the previous one only in the PML domain. Owing to the fact that Maxwell's equations have unique solutions, $\lim_{\omega \to \widetilde{\omega}_m}(\omega - \widetilde{\omega}_m)[\mathbf{E}_s(\mathbf{r}, \omega), \mathbf{H}_s(\mathbf{r}, \omega)] = \lim_{\omega \to \widetilde{\omega}_m}(\omega - \widetilde{\omega}_m)[\mathbf{E}'_s(\mathbf{r}, \omega), \mathbf{H}'_s(\mathbf{r}, \omega)]\ \forall\ \mathbf{r} \in \Omega \cap \Omega'$. Therefore, the $\gamma_m$ coefficient is unchanged and since the numerator in Eq. (B.3) is independent of the PML, the denominator, i.e., the PML norm, is shown to be also independent of the PML. QED.

**Appendix C. Non-reciprocal materials**

In this Appendix, we focus on the case where the reciprocity of the system is broken, namely, $\boldsymbol{\mu} \neq \boldsymbol{\mu}^T$ or $\boldsymbol{\varepsilon} \neq \boldsymbol{\varepsilon}^T$ and derive the orthonormalization condition for regularized QNMs.

Non-reciprocal materials are usually found in magneto-optical materials, such as yttrium iron garnet (YIG), iron garnets ($SrFe_{12}O_{19}$), biased by an external or static magnetic field to break the time-reversal symmetry [Asa20]. Optical resonators made of non-reciprocal materials are essential for many applications, such as one-way propagation, optical isolating, broadband field localization, and so on. To provide an intuitive understanding of these systems, it would be of interest to develop a method to normalize the QNMs when the reciprocity is broken.

The norm and orthogonality relationship for the QNMs of non-reciprocal resonators can be easily generalized from those of the reciprocal systems mentioned above. The generalization relies on the introduction of a left QNM, $\tilde{\mathbf{E}}_m^{(L)}$, which can be found by solving the source-free Maxwell's operator with transposed permittivity and permeability, $\boldsymbol{\varepsilon}^T$ and $\boldsymbol{\mu}^T$ [Zol18]. Plugging the left QNM and another QNM of the concerned system $\tilde{\mathbf{E}}_n^{(R)}$ (i.e., the mode of the system with $\boldsymbol{\varepsilon}$ and $\boldsymbol{\mu}$, we refer as the right QNM) into the Lorentz reciprocity formula, and noticing that they satisfy the source-free Maxwell's equations, we obtain the relationship

$$\iiint_{\Omega \cup \Omega_{\text{PML}}} \tilde{\mathbf{H}}_m^{(L)} \cdot [\widetilde{\omega}_n \boldsymbol{\mu}(\widetilde{\omega}_n) - \widetilde{\omega}_m \boldsymbol{\mu}(\widetilde{\omega}_m)] \cdot \tilde{\mathbf{H}}_n^{(R)} - \tilde{\mathbf{E}}_m^{(L)} \cdot [\widetilde{\omega}_n \boldsymbol{\varepsilon}(\widetilde{\omega}_n) - \widetilde{\omega}_m \boldsymbol{\varepsilon}(\widetilde{\omega}_m)] \cdot \tilde{\mathbf{E}}_n^{(R)} dV = 0, \quad (C1)$$

Note that, for nonreciprocal systems, $\widetilde{\mathbf{H}}_n^{(L)} \neq \widetilde{\mathbf{H}}_n^{(R)}$ and $\widetilde{\mathbf{E}}_n^{(L)} \neq \widetilde{\mathbf{E}}_n^{(R)}$, but they share the same eigenvalues, that is $\widetilde{\omega}_n^{(R)} = \widetilde{\omega}_n^{(L)} = \widetilde{\omega}_n$. The volume integral is performed over the whole space including the PML regions.

Further applying the Lorentz reciprocity to the left QNM and the scattered field of the resonator when it is driven by a dipole source emitting close to the resonance frequency, strictly following the approach in [Sau13] and using Eq. (C1), the QNM norm for non-reciprocal systems is derived

$$QN_n = \iiint_{\Omega \cup \Omega_{\text{PML}}} \widetilde{\mathbf{E}}_n^{(L)} \cdot \frac{\partial (\omega \boldsymbol{\varepsilon})}{\partial \omega} \cdot \widetilde{\mathbf{E}}_n^{(R)} - \widetilde{\mathbf{H}}_n^{(L)} \cdot \frac{\partial (\omega \boldsymbol{\mu})}{\partial \omega} \cdot \widetilde{\mathbf{H}}_n^{(R)} dV. \quad (C2)$$

For compactness, the detailed derivation of this formula is not shown here, and we refer the interested reader to Section 1.3.3 of [Wu21c].

The introduction of left QNMs in the normalization and orthogonality relations impacts the definition of some important physical quantities, compared to reciprocal cases. For example, the mode volume for non-reciprocal systems takes the following form [Wu21a]

$$\widetilde{V}_n(\mathbf{r}, \mathbf{u}) = \left[ 2\varepsilon_0 \left( \widetilde{\mathbf{E}}_n^{(L)} \cdot \mathbf{u} \right) \left( \widetilde{\mathbf{E}}_n^{(R)} \cdot \mathbf{u} \right) / QN_n \right]^{-1}, \quad (C3)$$

where $\mathbf{u}$ is the polarization direction.

Figure C1 shows the field and mode volume maps of a QNM of a 2D yttrium-iron garnet wire in a homogenous air background. Under an external dc magnetic field along the $z$ direction, the wire exhibits a strong gyromagnetic anisotropy, for a relative permeability tensor taking the form $\boldsymbol{\mu} = [\mu_r, -i k_r, 0; i k_r, \mu_r, 0; 0,0, \mu_\infty]$. In contrast to the reciprocal resonators, the fields for the left and right QNMs are different, as can be seen from Fig. C1(a). Therefore, to compute the mode volume in Fig. C1(b), two independent computations are required.

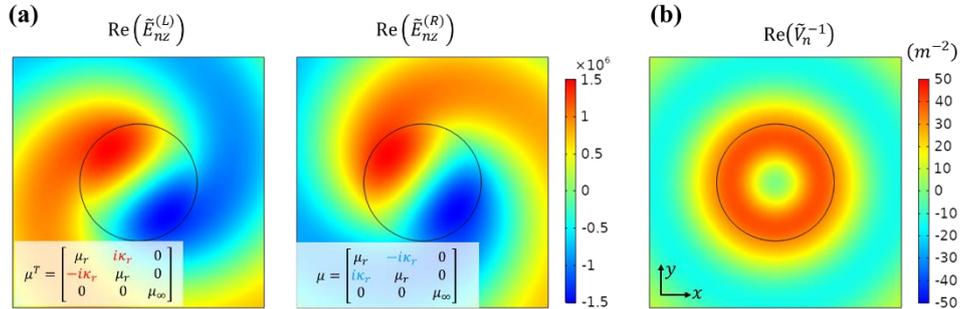

**Fig. C1.** (a) QNM field maps of a 2D yttrium-iron garnet wire in an air background. The first and second panels correspond to the maps for the real part of the electric field $z$-component for the left ($\widetilde{\mathbf{E}}_n^{(L)}$) and right ($\widetilde{\mathbf{E}}_n^{(R)}$) QNMs, respectively. The maps are computed for a QNM with its electric field parallel to the $z$ direction and an eigenfrequency of $7.90 + 0.04i$ GHz. The radius of the rod is 9.1 mm and it has an isotropic relative permittivity of $\varepsilon = 15$. (b) The map of $\text{Re}(\widetilde{V}_n^{-1})$ of the QNM with $\mathbf{u} = \mathbf{e}_z$. Note that because here we consider a 2D resonator, the mode volume takes the unit of $m^2$ instead of $m^3$ in 3D cases. This COMSOL model is freely available in the freeware package MAN [MAN].

## Appendix D. Discussion on the numerical implementation of the M norm

As discussed see the Section 4.1.2, several equivalent expressions of the M norm have been published over the years. The first expression involves only the electric field, with first and second-order spatial derivative [Mul10,Mul16a]. The M norm can also be formulated with only first-order spatial derivatives of the electric field, as shown in Appendix B of [Mul16a]. Finally, a more general expression valid for chiral and bi-anisotropic materials requires only the calculation of first-order spatial derivatives of the electric and the magnetic field, see Eq. (18) or [Mul18]. In the case of a sphere, the electromagnetic field is known analytically, and the spatial derivatives can be exactly calculated. For a different geometry, the QNM field, and thus its derivatives, must be calculated numerically. The numerical evaluation of the derivatives introduces a small numerical error in the surface integral. As a result, the divergence of the volume integral is not perfectly compensated, and a small error is introduced in the norm. The magnitude of the error depends on the implementation of the M norm. This issue has been investigated in Appendix G of [Mul16a].

Figure D1 displays a comparison between both implementations of the M norm if the derivatives are calculated numerically with a centered finite-difference scheme. The relative error between the PML norm and the M norm oscillates around zero with an amplitude that increases exponentially with $R$. An implementation of the M norm with only first-order derivatives (Eq. (19), dashed red curve) reduces the error by three orders of magnitude compared to the implementation with second-order derivatives (solid blue curve), see Fig. D1. The error could be further reduced by using a different numerical method to calculate the spatial derivative. However, methods more accurate than the centered finite-difference scheme usually increase the computational load since they require a more precise knowledge of the field behavior in the vicinity of the derivation point, i.e., multiple (larger than 2) numerical calculations of the field [Pre07]. Note that, for both implementations, the size of the domain $\Omega$ that provides the smallest error is in the range $R \approx \lambda - 2\lambda$. The use of smaller integration domains results in larger errors.

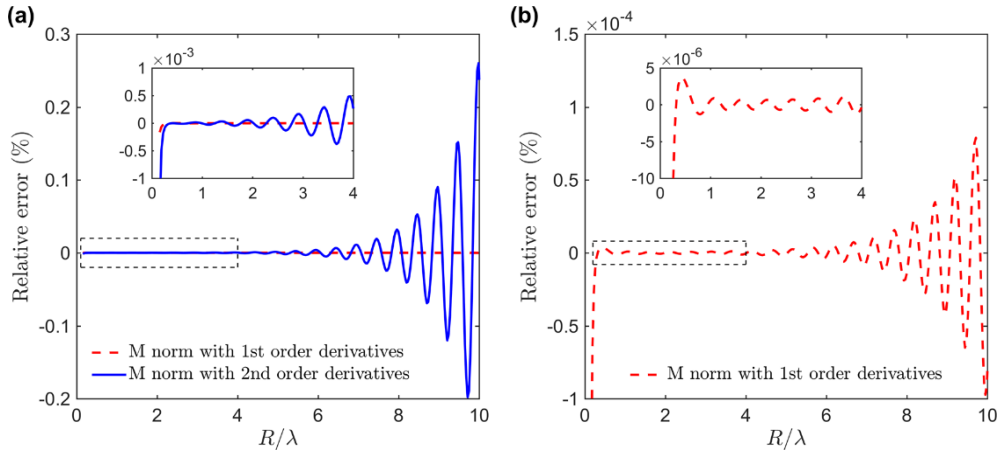

**Fig. D1.** Comparison between two different implementations of the M normalization for the same sphere as in Fig. 7. The spatial derivatives in the surface term $I_{\text{surf}}^{\text{M}}$ are calculated with a centered finite-difference scheme. We plot the same relative error as in Fig. 7. The solid blue curve shows the result of an implementation with second-order derivatives and electric field only (Eq. (8) in [Mul16a]) and the dashed red curve shows the result of an implementation with only first-order derivatives of the electric and magnetic fields contributing on equal footing, see Eqs. (18)-(19) and [Mul18]. The latter reduces the relative error by three orders of magnitude. (b) provides an enlarged view of (a). The insets display a zoom of the areas inside the rectangular boxes. Note that the dashed red curve is the same as in Fig. 7.

## Appendix E. Demonstration of Eq. (22)

In this Appendix, we demonstrate Eq. (22), i.e., a relation between the surface integral $I_{\text{surf}}^{\text{LK}}$ appearing in the LK norm and a volume integral of the electric and magnetic fields over a spherical domain $\Omega$. The expression of the surface integral $I_{\text{surf}}^{\text{LK}}$ is given by Eq. (17).

Let us consider a QNM ($\tilde{\mathbf{E}}_m, \tilde{\mathbf{H}}_m, \tilde{\omega}_m$) solution of Maxwell's equations in the absence of source. We first apply the divergence theorem to the vector $\tilde{\mathbf{E}}_m \times \tilde{\mathbf{H}}_m$ and the closed spherical surface $\Sigma$ defining the volume $\Omega$. Simple manipulations similar to the derivation of Lorentz reciprocity theorem in Appendix A lead to

$$\iint_\Sigma \tilde{\mathbf{E}}_m \times \tilde{\mathbf{H}}_m \cdot \mathbf{u}\,dS = i\tilde{\omega}_m \iiint_\Omega \left(\tilde{\mathbf{E}}_m \cdot \boldsymbol{\varepsilon}\tilde{\mathbf{E}}_m + \tilde{\mathbf{H}}_m \cdot \boldsymbol{\mu}\tilde{\mathbf{H}}_m\right)d^3\mathbf{r}, \tag{E1}$$

with **u** is the unit vector normal to the sphere $\Sigma$ (oriented outward).

For a resonator surrounded by a homogeneous medium and a large domain $\Omega$, one can use radiation conditions on the surface $\Sigma$ that provide a simple relation between the electric and magnetic fields. In the far field, the electromagnetic field locally behaves as an outgoing plane wave propagating along **u**. Therefore, for positions **r** on the surface $\Sigma$, one can write

$$\widetilde{\mathbf{H}}_m \approx \varepsilon_0 n c \mathbf{u} \times \widetilde{\mathbf{E}}_m, \tag{E2}$$

where $n$ is the refractive index of the surrounding medium and $c$ is the speed of light. The $\approx$ sign simply comes from the fact that Eq. (E2) is not rigorous for a finite-size domain $\Omega$. It is only asymptotically valid as $r \to \infty$.

Inserting Eq. (E2) into the left-hand side of Eq. (E1) and using the fact that the electric field of a plane wave is transverse to its propagation direction leads to

$$\iint_\Sigma \widetilde{\mathbf{E}}_m \times \widetilde{\mathbf{H}}_m \cdot \mathbf{u} \, dS \approx \varepsilon_0 n c \iint_\Sigma \widetilde{\mathbf{E}}_m^2 \, dS. \tag{E3}$$

According to the definition of the surface term $I_{\text{surf}}^{\text{LK}}$ in Eq. (17), Eq. (E1) simply becomes

$$I_{\text{surf}}^{\text{LK}} \approx -\iiint_\Omega \left(\widetilde{\mathbf{E}}_m \cdot \boldsymbol{\varepsilon} \widetilde{\mathbf{E}}_m + \widetilde{\mathbf{H}}_m \cdot \boldsymbol{\mu} \widetilde{\mathbf{H}}_m\right) d^3\mathbf{r}. \tag{E4}$$

## 7. References

[Abd18] M. I. Abdelrahman and B. Gralak, "Completeness and divergence-free behavior of the quasi-normal modes using causality principle," OSA Continuum **1**, 340-348 (2018).
[Aks14] G. M. Akselrod, C. Argyropoulos, T. B. Hoang, C. Ciracì, C. Fang, J. Huang, D. R. Smith, and M. H. Mikkelsen, "Probing the mechanisms of large Purcell enhancement in plasmonic nanoantennas," Nat. Photonics **8**, 835–840 (2014).
[Arc10] A. Archambault, F. Marquier, J.-J. Greffet, and C. Arnold, "Quantum theory of spontaneous and stimulated emission of surface plasmons," Phys. Rev. B **82**, 035411 (2010).
[Arf05] G. B. Arfken and H. J. Weber, *Mathematical methods for physicists*, 6th edition, chapter 7, (Elsevier Academic Press, Amsterdam, 2005).
[Asa20] V. S. Asadchy, M. S. Mirmoosa, A. Díaz-Rubio, S. Fan, and S. A. Tretyakov, "Tutorial on electromagnetic nonreciprocity and its origins," S. A. Proceedings of the IEEE **108**, 1684-1727 (2020).
[Ash20] Y. Ashida, Z. Gong, and M. Ueda, "Non-Hermitian Physics," Adv. In Physics **69**, 249-435 (2020).
[Bai13] Q. Bai, M. Perrin, C. Sauvan, J-P Hugonin, and P. Lalanne, "Efficient and intuitive method for the analysis of light scattering by a resonant nanostructure," Opt. Express **21**, 27371 (2013).
[Bau71] C. E. Baum, Interaction Note 88, December 1971. See https://apps.dtic.mil/sti/citations/ADA066905
[Bau76] C. E. Baum, "The singularity expansion method," p. 129-179 in *Transient Electromagnetic Fields* Springer Berlin Heidelberg, L. B. Felsen ed. (1976).
[Baz69] A. I. Baz, Ya. B. Zel'dovich, and A. M. Perelomov, Scattering, Reactions, and Decay in Non-Relativistic Quantum Mechanics (IPST, 1969) (Moskva, Nauka, 1971).
[Ber07] J. P. Berenger, Perfectly Matched Layer (PML) for Computational Electromagnetic (Morgan & Claypool, 2007).
[Bet21] F. Betz, F. Binkowski, and S. Burger, "RPExpand: Software for Riesz projection expansion of resonance phenomena," SoftwareX **15**, 100763 (2021).
[Bru16] Y. Brûlé, B. Gralak, and G. Demésy, "Calculation and analysis of the complex band structure of dispersive and dissipative two-dimensional photonic crystals," J. Opt. Soc. Am. B **33**, 691 (2016).
[Bur21] S. Burger, personal communication (2021).
[Che94] W. C. Chew and W. H. Weedon, "A 3D perfectly matched medium from modified Maxwell's equations with stretched coordinates," Microwave Opt. Technol. Lett. **7**, 599–604 (1994).

[Chr19] T. Christopoulos, O. Tsilipakos, G. Sinatkas, and E. E. Kriezis, "On the calculation of the quality factor in contemporary photonic resonant structures," Opt. Express **27**, 14505-22 (2019).

[Cog20] K. G. Cognée, *Hybridization of open photonic resonators*, PhD dissertation, University of Bordeaux and University of Amsterdam (2020).

[Col18] R. Colom, R. McPhedran, B. Stout, and N. Bonod, "Modal expansion of the scattered field: Causality, nondivergence, and nonresonant contribution," Phys. Rev. B **98**, 085418 (2018).

[Dem20] G. Demésy, A. Nicolet, B. Gralak, C. Geuzaine, C. Campos, and J. E. Roman, "Non-linear eigenvalue problems with GetDP and SLEPc: Eigenmode computations of frequency-dispersive photonic open structures," Comput. Phys. Commun. **257**, 107509 (2020).

[Dez18] M. K. Dezfouli and S. Hughes, "Regularized quasinormal modes for plasmonic resonators and open cavities," Phys. Rev. B **97**, 115302 (2018).

[Doo13] M. B. Doost, W. Langbein, and E. A. Muljarov, "Resonant state expansion applied to two-dimensional open optical systems," Phys. Rev. A **87**, 043827 (2013).

[Doo14] M. B. Doost, W. Langbein, and E. A. Muljarov, "Resonant-state expansion applied to three-dimensional open optical systems," Phys. Rev. A **90**, 013834 (2014).

[El18] R. El-Ganainy, K. G. Makris, M. Khajavikhan, Z. H. Musslimani, S. Rotter, and D. N. Christodoulides, "Non-Hermitian physics and PT symmetry," Nat. Phys. **14**, 11-19 (2018).

[Fag17] R. Faggiani, J. Yang, R. Hostein, and P. Lalanne, "Implementing structural slow light on short length scales: the photonic speed-bump," Optica **4**, 393-399 (2017). The normalization of cavities coupled to periodic waveguides has also be studied in Opt. Lett. **39**, 6359 (2014) by regularizing the divergent field in the waveguide with advanced theorems in complex analysis. It would be interesting to compare the two approaches.

[Fah09] S. Fahr, C. Rockstuhl, and F. Lederer, "Metallic nanoparticles as intermediate reflectors in tandem solar cells," Appl. Phys. Lett. **95**, 121105 (2009).

[Fei82] P. J. Feibelman, "Surface electromagnetic fields," Prog. Surf. Sci. **12**, 287-407 (1982).

[Fra19] S. Franke, S. Hughes, M. K. Dezfouli, P. T. Kristensen, K.Busch, A. Knorr, and M. Richter, "Quantization of Quasinormal Modes for Open Cavities and Plasmonic Cavity Quantum Electrodynamics," Phys. Rev. Lett. **122**, 213901 (2019).

[Fra20] S. Franke, M. Richter, J. Ren, A. Knorr, and S. Hughes, "Quantized quasinormal mode description of non-linear cavity QED effects from coupled resonators with a Fano-like resonance," Phys. Rev. Research **2**, 033456 (2020).

[Gam28] G. Gamow, "Quantum Theory of the Atomic Nucleus," G. Z. Physik **51**, 204 (1928).

[Ge14] R.-C. Ge, P. T. Kristensen, J. F Young, and S. Hughes, "Quasinormal mode approach to modelling light-emission and propagation in nanoplasmonics," New J. Phys. **16**, 113048 (2014).

[Gra19] A. Gras, W. Yan, and P. Lalanne, "Quasinormal-mode analysis of grating spectra at fixed incidence angles," Opt. Lett. **44**, 3494-3497 (2019).

[Gra20] A. Gras, P. Lalanne, and M. Duruflé, "Non-uniqueness of the Quasinormal Mode Expansion of Electromagnetic Lorentz Dispersive Materials," J. Opt. Soc. Am. A **37**, 1219-28 (2020).

[Har61] R. F. Harrington, *Time Harmonic Electromagnetic Fields* (McGraw-Hill, New York, 1961).

[Hei04] S. Hein, T. Hohage, and W. Koch, "On resonances in open systems," J. Fluid. Mech. **506**, 255-284 (2004).

[Hoe79] B. J. Hoenders, "On the completeness of the natural modes for quantum mechanical potential scattering," J. Math. Phys. **20**, 329-335 (1979).

[Hug05] J.-P. Hugonin and P. Lalanne, "Perfectly-matched-layers as nonlinear coordinate transforms: a generalized formalization," J. Opt. Soc. Am. A **22**, 1844-1849 (2005).

[Jar21] J. L. Jaramillo, R. P. Macedo, and L. Al Sheikh, "Pseudospectrum and Black Hole Quasinormal Mode Instability," Phys. Rev. X **11**, 031003 (2021).

[Joa08] J. D. Joannopoulos, S. G. Johnson, J. N. Winn, and R. D. Meade, *Photonic Crystals: Molding the Flox of Light*, 2nd edition (Princeton University Press, Princeton, 2008).

[Kip08] T. J. Kippenberg and K. J. Vahala, "Cavity Optomechanics: Back-Action at the Mesoscale," Science **321**, 1172 (2008).

[Kok99] K. D. Kokkotas and B. G. Schmidt, "Quasi-normal modes of stars and black holes," Living Rev. Relativ. **2**, 2 (1999).
[Kon11] R.A. Konoplya and A. Zhidenko, "Quasinormal modes of black holes: From astrophysics to string theory," Rev. Mod. Phys. **83**, 793-836 (2011).
[Kri12] P. T. Kristensen, C. Van Vlack, and S. Hughes, "Generalized effective mode volume for leaky optical cavity," Opt. Lett. **37**, 1649 (2012).
[Kri14a] P. T. Kristensen, J. R. de Lasson, and N. Gregersen, "Calculation, normalization and perturbation of quasinormal modes in coupled cavity-waveguide systems," Opt. Lett., 39, 6359-6362 (2014).
[Kri14b] P.T. Kristensen and S. Hughes, "Modes and Mode Volumes of Leaky Optical Cavities and Plasmonic Nanoresonators," ACS Photonics **1**, 2-10 (2014).
[Kri15] P. T. Kristensen, R.-C. Ge, and S. Hughes, "Normalization of quasinormal modes in leaky optical cavities and plasmonic resonators," Phys. Rev. A **92**, 053810 (2015).
[Kri17] P.T. Kristensen, R.C. Ge, and S. Hughes, "Reply to "Comment on 'Normalization of quasinormal modes in leaky optical cavities and plasmonic resonators'," Phys. Rev. A **96**, 017802 (2017).
[Kri20] P. T. Kristensen, K. Herrmann, F. Intravaia, and K. Busch, "Modeling electromagnetic resonators using quasinormal modes," Adv. Opt. Photon. **12**, 612-708 (2020).
[Lai90] H. M. Lai, P. T. Leung, K. Young, P. W. Barber, and S. C. Hill, "Time-independent perturbation for leaking electromagnetic modes in open systems with application to resonances in microdroplets," Phys. Rev. A **41**, 5187 (1990).
[Lal08] P. Lalanne, C. Sauvan, and J.-P. Hugonin, "Photon confinement in photonic crystal nanocavities," Laser Photonics Rev. **2**, 514 (2008).
[Lal18] P. Lalanne, W. Yan, K. Vynck, C. Sauvan, and J.-P. Hugonin, "Light interaction with photonic and plasmonic resonances," Laser Photonics Rev. **12**, 1700113 (2018).
[Lal19] P. Lalanne, W. Yan, A. Gras, C. Sauvan, J.-P. Hugonin, M. Besbes, G. Demesy, M. D. Truong, B. Gralak, F. Zolla, A. Nicolet, F. Binkowski, L. Zschiedrich, S. Burger, J. Zimmerling, R. Remis, P. Urbach, H. T. Liu, and T. Weiss, "Quasinormal mode solvers for resonators with dispersive materials," J. Opt. Soc. Am. A **36**, 686-704 (2019).
[Lal20] P. Lalanne, "Mode volume of electromagnetic resonators: let us try giving credit where it is due," arXiv:2011.00218 (2020).
[Lee09] S.-Y. Lee, "Decaying and growing eigenmodes in open quantum systems: Biorthogonality and the Petermann factor," Phys. Rev. A **80**, 042104 (2009).
[Lee99] K. M. Lee, P.T. Leung, and K.M. Pang, "Dyadic formulation of morphology-dependent resonances. I. Completeness relation," J. Opt. Soc. Am. B **16**, 1409-1417 (1999).
[Leo06] U. Leonhardt and T. G. Philbin, "General relativity in electrical engineering," New J. Phys. **8**, 247 (2006).
[Leu94a] P. T. Leung, S. Y. Liu, and K. Young, "Completeness and orthogonality of quasinormal modes in leaky optical cavities," Phys. Rev. A **49**, 3057 (1994).
[Leu94b] P. T. Leung, S. Y. Liu, and K. Young, "Completeness and time-independent perturbation of the quasinormal modes of an absorptive and leaky cavity," Phys. Rev. A **49**, 3982 (1994).
[Leu96] P. T. Leung and K. M. Pang, "Completeness and time-independent perturbation of morphology-dependent resonances in dielectric spheres," J. Opt. Soc. Am. B **13**, 805-817 (1996).
[MAN] MAN (Modal Analysis of Nanoresonators). QNM solvers and toolboxes are freely available at https://www.lp2n.institutoptique.fr/light-complex-nanostructures
[Man17] M. Mansuripur, M. Kolesik, and P. Jakobsen, "Leaky modes of solid dielectric spheres," Phys. Rev. A **96**, 013846 (2017).
[Mar91] D. Marcuse, Theory of Dielectric Optical Waveguides, 2nd ed. (Academic, New York, 1991).
[Moi11] N. Moiseyev, *Non-Hermitian quantum mechanics*, (Cambridge University Press, 2011).
[Moi98] N. Moiseyev, "Quantum theory of resonances: calculating energies, widths and cross-sections by complex scaling," Phys. Rep. **302**, 211-293 (1998).
[Mor71] R. M. More, "Theory of decaying states," Phys. Rev. A **4**, 1782 (1971).

[Mor73] R. M. More and E. Gerjuoy, "Properties of Resonance Wave Functions," Phys. Rev. A **7**, 1288 (1973).
[Mul10] E. A. Muljarov, W. Langbein, and R. Zimmermann, "Brillouin-Wigner perturbation theory in open electromagnetic systems," Europhys. Lett. **92**, 50010 (2010).
[Mul16a] E. A. Muljarov and W. Langbein, "Exact mode volume and Purcell factor of open optical systems," Phys. Rev. B **94**, 235438 (2016).
[Mul16b] E. A. Muljarov and W. Langbein, "Resonant-state expansion of dispersive open optical systems: Creating gold from sand," Phys. Rev. B **93**, 075417 (2016).
[Mul17] E. A. Muljarov and W. Langbein, "Comment on 'Normalization of quasinormal modes in leaky optical cavities and plasmonic resonators'," Phys. Rev. A **96**, 017801 (2017).
[Mul18] E. A. Muljarov and T. Weiss, "Resonant-state expansion for open optical systems: generalization to magnetic, chiral, and bi-anisotropic materials," Opt. Lett. **43**, 1978 (2018).
[Nic94] A. Nicolet, J.-F. Remacle, B. Meys, A. Genon, and W. Legros, "Transformation methods in computational electromagnetics," J. Appl. Phys. **75**, 6036-6038 (1994).

[note1] Exceptions occurs when one tries to lower substrate reflection by choosing the superstrate medium to have the same refractive index as the substrate.
[note2] In 2D, even for the ideal case of resonators in uniform backgrounds, there is always a singularity in 2D owing to the singularity of the 2D Hankel function at the origin $r = 0$.
[note3] The term 'PML modes' or 'Bérenger modes' is particularly used to describe the modes that are obtained with finite-thickness PMLs and are intended to replace the continuous spectrum of the operator resulting from the coupling with the open space. However, the PML modes are also obtained for discretized operator in practice. Discretization always induces errors especially at high frequency owing to the necessarily finite mesh elements. In addition, with non-dispersive PMLs that are quite often used, numerical errors might also occur at low frequencies. Therefore, the ideal PML modes are in practice contaminated by the unperfect discretization. To account for that point, we prefer to call the 'PML modes' 'numerical modes' in the review.
[note4] The unicity of the $\alpha$'s singularly contrasts with the case of expansions based on pole expansions, for which there is an infinity of mathematically-correct formulas owing to the fact that the QNM basis is overcomplete [Pea81], see the Eqs. (2.23) and (2.24) in [Leu94a] and related discussions in [Mul16b,Doo13,Doo14].
[note5] Our presentation remains simplistic, and the PML concept is not fully general. It may completely fail for instance for geometries that are not spatially invariant, e.g., half spaces with periodic waveguides. Even translation invariant half spaces supporting backward waves with opposite phase and group velocities may lead to insurmountable difficulties, see for instance E. Bécache, S. Fauqueux and P. Joly, "Stability of perfectly matched layers, group velocities and anisotropic waves", Journal of Computational Physics **188**, 399-433 (2003).
[note6] Note that $x$' if often denoted simply $x$, which leads to a confusion since this $x$ is different from the original $x$ coordinate in Eq. (1). In practice, in numerics, $x$' corresponds to the numerical space inside the PML.
[note7] The literature has taken the habit to distinguish two kinds of modes, the QNMs and the PML modes. To lift it and also to be more general and encompass methods that do not rely on PMLs, e.g., integral methods [see D. A. Powell, "Interference between the Modes of an All-Dielectric Meta-atom," Phys. Rev. Appl. **7**, 034006 (2017)], hereafter we will call the PML modes 'numerical modes'. The latter are affected by the boundary condition (e.g., hard wall for finite PMLs), whereas the former is not.
[note8] According to the experience of some of the authors, using dispersive PMLs makes the numerical implementation for computing the QNMs more difficult but does not change the quality of the reconstruction.
[note9] The demonstration in [Yan18] relies on recognizing that the permittivity of any linear dispersive material can be modeled with a *N*-pole Lorentz-pole permittivity, with increasing accuracy as the number of poles N increases.

[note10] The term biorthogonal (rather than orthogonal) product is used because the product involves right and left eigenvectors, which are identical for reciprocal systems, see Section3.4.3 and Appendix C, where the normalization of QNMs or resonators with non-reciprocal materials is considered.

[note11] By most matrices, we intend to mean that the set of diagonalizable matrices (in fact, even the set of matrices with distinct eigenvalues), is a set that is dense in the algebra of matrices with complex coefficients.

[note12] The existence of an infinity of expressions for the excitation coefficient should not be confused with the so called 'overcompleteness' of QNM expansions inside resonators places in free space. Overcompleteness is due to the fact that only a compact subspace (the resonator inside) of the whole open space is used for the expansion. With regularization, the expansion holds over the whole space and the infinity of expressions arises from the infinity of auxiliary fields for dispersive materials only.

[note13] In fact, completeness is likely to be achieved in the whole computational domain, even in the PML regions.

[note14] There exist many formulas for $\alpha_m(\omega)$. However, there is no ambiguity since all the formulas should gives the same value as $\omega$ tends towards the resonance frequency $\widetilde{\omega}_m$, see Eq. (4.8) in [Lal18].

[Oly04] F. Olyslager, "Discretization of Continuous Spectra Based on Perfectly Matched Layers," SIAM J. Appl. Math. **64**, 1408-1433 (2004).

[Par20] M. Parto, Y. G. N. Liu, B. Bahari, M. Khajavikhan, and D. N. Christodoulides, "Non-Hermitian and topological photonics: optics at an exceptional point," Nanophotonics **10**, 403-423 (2020).

[Pea81] L. W. Pearson, "Evidence that bears on the left half plane asymptotic behavior of the SEM expansion of surface currents," Electromagnetics **1**, 395-402 (1981).

[Pre07] W. H. Press, S. A. Teukolsky, W. T. Vetterling and, B. P. Flannery, [Numerical Recipes], (Cambridge University Press, New York 2007).

[Pre71] W. H. Press, "Long wave trains of gravitational waves from a vibrating black hole," Astrophys. J. **170**, L105-L108 (1971).

[QNMPole] QNMPole is one of the two QNM solvers of the freeware MAN (Modal Analysis of Nanoresonators) freely available at https://www.lp2n.institutoptique.fr/light-complex-nanostructures.

[Ram10] A. Raman and S. Fan, "Photonic Band Structure of Dispersive Metamaterials Formulated as a Hermitian Eigenvalue Problem," Phys. Rev. Lett. **104**, 087401 (2010).

[Riv20] N. Rivera and I. Kaminer, "Light-matter interactions with photonic quasiparticles," Nat. Rev. Phys.**2**, 538-561 (2020).

[Röm08] F. Römer and B. Witzigmann, "Spectral and spatial properties of the spontaneous emission enhancement in photonic crystal cavities," J. Opt. Soc. Am. B **25**, 31-39 (2008).

[San16] K. Santosh, O. Bitton, L. Chuntonov, and G. Haran, "Vacuum Rabi splitting in a plasmonic cavity at the single quantum emitter limit," Nat. Commun. **7**, 11823 (2016).

[Sau13] C. Sauvan, J.P. Hugonin, I.S. Maksymov, and P. Lalanne, "Theory of the spontaneous optical emission of nanosize photonic and plasmon resonators," Phys. Rev. Lett **110**, 237401 (2013).

[Sau15] C. Sauvan, J.-P. Hugonin, and P. Lalanne, "Photonic and plasmonic nanoresonators: a modal approach," Proc. SPIE, Active Photonic Materials VII **9546**, 95461C (2015). doi: 10.1117/12.2190201

[Sau21] C. Sauvan, "Quasinormal modes expansions for nanoresonators made of absorbing dielectric materials: study of the role of static modes," Opt. Express **29**, 8268 (2021).

[Set03] A. Settimi, S. Severini, N. Mattiucci, C. Sibilia, M. Centini, G. D'Aguanno, and M. Bertolotti, "Quasinormal-mode description of waves in one-dimensional photonic crystals," Phys. Rev. E **68**, 026614 (2003).

[Set09] A. Settimi, S. Severini, and B. J. Hoenders, "Quasi-normal-modes description of transmission properties for photonic bandgap structures," J. Opt. Soc. Am. B **26**, 876 (2009).

[Sha19] A. M. Shaltout, V. M. Shalaev, and M. L. Brongersma, "Spatiotemporal light control with active metasurfaces," Science **364**, eaat3100 (2019).

[Sie39] A. J. F. Siegert, "On the Derivation of the Dispersion Formula for Nuclear Reactions," Phys. Rev. **56**, 750-752 (1939).

[Sil01] E. Silberstein, P. Lalanne, J.P. Hugonin, and Q. Cao, "On the use of grating theory in integrated optics," J. Opt. Soc. Am. A. **18**, 2865-2875 (2001).
[Sny83] A. W. Snyder and J. D. Love, *Optical Waveguide Theory* (Chapman and Hall, London, 1983).
[Sto21] B. Stout, R. Colom, N. Bonod, and R. C. McPhedran, "Spectral expansions of open and dispersive optical systems: Gaussian regularization and convergence," New J. Phys. **23**, 083004 (2021).
[Tan07] T. Tanabe, M. Notomi, E. Kuramochi, A. Shinya, and H. Taniyama, "Trapping and delaying photons for one nanosecond in an ultrasmall high-Q photonic-crystal nanocavity," Nat. Photon. **1**, 49 (2007).
[Tao20] C. Tao, J. Zhu, Y. Zhong, and H. Liu, "Coupling theory of quasinormal modes for lossy and dispersive plasmonic nanoresonators," Phys. Rev. B **102**, 045430 (2020).
[Tru20] M. D. Truong, A. Nicolet, G. Demésy, and F. Zolla, "Continuous family of exact Dispersive Quasi-Normal Modal (DQNM) expansions for dispersive photonic structures," Opt. Express **28**, 29016-29032 (2020).
[Tur06] H. E. Türeci, A. D. Stone, and B. Collier, "Self-consistent multimode lasing theory for complex or random lasing media," Phys. Rev. A **74**, 043822 (2006).
[Via14] B. Vial, A. Nicolet, F. Zolla, and M. Commandré, "Quasimodal expansion of electromagnetic fields in open two-dimensional structures," Phys. Rev. A **89**, 023829 (2014).
[Wei18] T. Weiss and E. A. Muljarov, "How to calculate the pole expansion of the optical scattering matrix from the resonant states," Phys. Rev. B, **98**, 085433 (2018).
[Wu21a] T. Wu, M. Gurioli, and P. Lalanne, "Nanoscale Light Confinement: the Q's and V's," ACS Photonics **8**, 1522-1538 (2021).
[Wu21b] T. Wu, D. Arrivault, M. Duruflé, A. Gras, F. Binkowski, S. Burger, W. Yan, and P. Lalanne, "Efficient hybrid method for the modal analysis of optical microcavities and nanoresonators," J. Opt. Soc. Am. A **38** 1224-1231 (2021)
[Wu21c] T. Wu and P. Lalanne, "QNMnonreciprocal_resonators: an openly available toolbox for computing the QuasiNormal Modes of nonreciprocal resonators," arXiv:2106.05502 (2021).
[Yan18] W. Yan, R. Faggiani, and P. Lalanne, "Rigorous modal analysis of plasmonic nanoresonators," Phys. Rev. B **97**, 205422 (2018).
[Yan20] W. Yan, P. Lalanne, and M. Qiu, "Shape Deformation of Nanoresonator: A Quasinormal-Mode Perturbation Theory," Phys. Rev. Lett. **125**, 013901 (2020).
[Zel61] Y. B. Zel'dovich, "On the theory of unstable states," Sov. Phys. JETP **12**, 542 (1961).
[Zha20] H. Zhang and O. D. Miller, "Quasinormal Coupled Mode Theory," arXiv:2010.08650 (2020).
[Zho21] Q. Zhou, P. Zhang, and X. Chen, "Quantum surface response informed quasinormal mode theory for nanoscale electromagnetism," arXiv:2105.06328 (2021).
[Zol18] F. Zolla, A. Nicolet, and G. Demésy, "Photonics in highly dispersive media: the exact modal expansion," Opt. Lett. **43**, 5813-5816 (2018).
[Zsc18] L. Zschiedrich, F. Binkowski, N. Nikolay, O. Benson, G. Kewes, and S. Burger, "Riesz-projection-based theory of light-matter interaction in dispersive nanoresonators," Phys. Rev. A **98**, 043806 (2018).